\documentclass[twocolumn]{aastex7}
\usepackage{natbib}
\usepackage{color}
\usepackage{booktabs}
\usepackage{hyperref} 
\usepackage{mathtools}
\usepackage{overpic}

\usepackage{comment}

\DeclarePairedDelimiterX\braket[2]{\langle}{\rangle}{#1\,\delimsize\vert\,\mathopen{}#2}

\def    \apjl  		{\rm {ApJL}}

\def    \apjl  		{\rm {ApJL}}

\def	\cm		{\,{\rm {cm}}}
\def	\K		{\,{\rm K}}
\def	\g		{\,{\rm {g}}}
\def	\mum	{\,{\mu \rm{m}}}

\def \bea {\begin{eqnarray}}
\def \ena {\end{eqnarray}}

\def	\bJ	{\boldsymbol{J}}

\def    \bmu    {{\hbox{\boldsym\char'026}}}	

\def	\cm	{\,{\rm cm}}

\def	\max	{\,{\rm max}}
\def	\d	{{\rm d}}

\def	\D	{{\rm D}}
\def	\eff	{{\rm eff}}

\def	\erg	{\,{\rm erg}}

\def	\g	{\,{\rm g}}
\def	\gas	{\,{\rm gas}}
\def	\G	{{\rm G}}

\def	\H	{{\rm H}}

\def	\N	{{\rm N}}

\def	\s	{\,{\rm s}}

\def	\d 	{\rm d}
\def	\B 	{\rm B}

\def \St {{\rm St}}

\def	\V	{{\rm V}}

\def	\yr	{{\rm yr}}

\def    \gas     	{{\rm gas}}

\font\mib=cmmib10

\def\bOmega{\hbox{\mib\char"0A}}
\def\bmu{\hbox{\mib\char"16}}

\makeatletter
\newcommand*{\rom}[1]{\expandafter\@slowromancap\romannumeral #1@}

\begin{document}

\title{Effective Magnetic Susceptibility of Dust Grains with Superparamagnetic Inclusions and Implications}

\author{Thiem Hoang}
\affiliation{Korea Astronomy and Space Science Institute, Daejeon 34055, Republic of Korea}
\affiliation{Department of Astronomy and Space Science, University of Science and Technology, 217 Gajeong-ro, Yuseong-gu, Daejeon, 34113, Republic of Korea}

\email{thiemhoang@kasi.re.kr}  
\date{Draft version \today} 

\begin{abstract}

Magnetic properties of dust play a fundamental role in grain alignment, dust polarization, and magnetic dipole emission. In the radiative torque (RAT) paradigm, superparamagnetic inclusions (SPIs) embedded within dust grains are expected to significantly enhance magnetic susceptibility and alignment efficiency. 
We introduce an effective superparamagnetism model that explicitly accounts for a size distribution of SPIs. We show that the effective zero-frequency susceptibility can be described by the superparamagnetic susceptibility of uniform-sized inclusions evaluated at the critical blocking size, reduced by a factor $F_{\rm eff}\sim 0.1$. For a fiducial SPI size-distribution slope of $q=11/6$, the effective susceptibility increases slightly with dust temperature $T_d$ until all SPIs become thermally active at $T_{d}\sim 110$ K, which contrasts with the $T_{d}^{-1}$ decline found for single-size SPIs. For rotating grains at angular frequency $\omega$, we identify a characteristic resonance size of SPIs that dominates the magnetic response, $N_{\rm res} = (T_{d}/T_{\rm act}) \ln (\nu_{0}/\omega)$ with $T_{\rm act}$ activation temperature and $\nu_{0}$ the characteristic attempt frequency of SPIs. The frequency-dependent effective susceptibility is well described by the zero-frequency susceptibility of uniform-sized SPIs at $N_{\rm res}$, reduced by a factor $G_{\rm eff}\sim 0.02$--$0.04$. Unlike models assuming uniform-sized inclusions, the effective susceptibility exhibits a nearly flat spectrum for frequency below $\nu_{0}$, arising from the progressive activation of larger inclusions at lower frequencies. For a narrow log-normal SPI size distribution, the response instead resembles a broadened single-size resonance set by the peak of the iron mass distribution, whereas a broad log-normal distribution behaves like the power law. 

\end{abstract}


\section{Introduction}
The alignment of non-spherical dust grains with ambient magnetic fields causes the polarization of distant starlight \citep{Hall.1949,Hiltner.1949} and polarized thermal dust emission \citep{Hildebrand:1988p2566,Planck.2015}. Dust polarization has become one of the primariy tools for tracing the magnetic fields (B-fields) projected in the plane of the sky (2D B-fields) across a wide range of astrophysical environments, from the diffuse interstellar medium (ISM) and molecular clouds to star- and planet-forming regions \citep{Pattle.2023ASPC,Tsukamoto.2023}. The polarization degree also provides insight into dust properties (size, shape, and composition) \citep{draine2024sensitivity,HoangBao.2024,TruongHoang.2025}, and the physical processes responsible for grain alignment \citep{Lazrev.2003,LAH.2015}. 

Magnetic properties of dust grains have been recognized as essential for grain alignment with magnetic fields since the earliest theoretical studies of interstellar dust alignment so-called paramagnetic relaxation mechanism \citep{DavisGreenstein.1951}. Iron, the fifth abundant element in the universe, is heavily depleted from the gas phase, with more than $95\%$ of iron incorporated into dust grains \citep{Jenkins.2009}. While the precise chemical and structural form of iron in dust remains uncertain, in-situ measurements and astronomical modeling suggest that a significant fraction of iron may be present in the form of metallic or iron-oxide clusters embedded within larger silicate grains. \cite{Bradley:1994p6379} first discovered the existence of iron inclusions with radius between 1-10 nm embedded within interplanetary silicate grains, aka. GEMs (glass with embedded metal and sulfides), which originate from interstellar grains (see \citealt{bradley2022provenance} for a review). Modern interstellar dust models, including THEMIS \citep{Kohler.2014,Jones.2017,Ysard.2024} and Astrodust \citep{Hensley.2023}, incorporate iron inclusions to reproduce observed extinction, emission, and polarization properties.

Incorporation of iron clusters of nanometer-size (hereafter superparamagnetic inclusions--SPIs) into a dust grain can strongly enhance the magnetic response of dust grains, making the composite grain superparamagnetic material. Early works noted that SPIs increase magnetic relaxation rates of rotating grains and thus induce more efficient alignment with magnetic fields \citep{JonesSpitzer.1967}, which was previously suggested as the dominant factor for interstellar grain alignment \citep{Mathis.1986,Martin.1995,Goodman.1995b}. In the modern alignment theory, namely, Radiative Torque (RAT) paradigm \citep{Dolginov:1976p2480,DraineWein.1997,LazHoang.2007,HoangLaz:2008gb}, grain alignment with ambient magnetic fields depends on three key ingredients: dust properties (including magnetic susceptibility), the radiation field, and local gas conditions (e.g., \citealt{Hoangetal.2022}). Superparamagnetism plays a particularly important role in Magnetically enhanced RAT (MRAT) alignment, where enhanced magnetic relaxation stabilizes the high-angular-momentum (high-J) attractor point and increases the fractions of grains aligned at high-J attractors, leading to higher alignment efficiency \citep{LazHoang.2008,HoangLaz.2016,LazHoang.2019,Hoang.2025}. Radiative torques can also induce rotational disruption of dust grains (aka. RATD) due to centrifugal stress when the grain angular velocity exceeds the critical limit determined by the material tensile strength \citep{Hoangetal.2019,Hoang.2019}. The RATD is an important mechanism that affects the evolution of dust in strong radiation fields, and its efficiency also depends on magnetic properties \citep{Hoang.2025}.

In addition to alignment, thermal fluctuations of magnetic moments in SPIs produce magnetic dipole emission (MDE), which has been proposed as a potential foreground component of the cosmic microwave background (CMB) \citep{DraineLaz.1999,HoangLaz.2016b}. Thus, the magnetic susceptibility of dust grains containing SPIs is fundamental not only for grain alignment but also for MDE.

Despite this importance, previous treatments of superparamagnetism in dust grains have generally assumed SPIs of a single characteristic size when computing magnetic susceptibility and relaxation rates. This simplifying assumption overlooks a crucial physical property: the superparamagnetic relaxation time depends exponentially on inclusion volume. In realistic astrophysical environments, iron clusters are expected to form with a range of sizes. Laboratory experiments demonstrate the formation of iron clusters spanning nanometer scales \citep{Davoisne:2006fn,Djouadi:2007p6409}, and theoretical studies suggest that the maximum size of SPIs depends on dust temperature and thermal activation \citep{Yang.2021}. Because the magnetic response is highly size-sensitive, a distribution of inclusion sizes can fundamentally alter the effective magnetic susceptibility of dust grains.

The size dependence becomes even more critical when considering rotating grains. For a grain rotating at angular frequency $\omega$, the magnetic response is frequency-dependent, and only SPIs with N\'eel relaxation times, which is determined by dust temperature, comparable to $\omega^{-1}$ contribute efficiently to magnetic response. As a result, the dominant inclusion size depends on both rotation rate and dust temperature. This frequency- and temperature-selective behavior cannot be captured by single-size SPI models. Moreover, while ordinary paramagnetism predicts a Curie-like decrease of static susceptibility with increasing temperature, the activation of larger SPIs at higher temperatures can modify this trend, potentially leading to qualitatively different temperature dependence of the effective susceptibility.

These considerations are particularly relevant in environments where dust is exposed to strong radiation fields in which not only dust temperature but grain angular momentum spunup by RATs can be increased. High-spatial-resolution polarization observations now can probe warm/hot dust in photodissociation regions, e.g., by SOFIA/HWAC+ \citep{TramHoang:2021,Thuong.2022,Ngoc.2024apj}, active galactic nuclei \citep{Lopez-Rodriguez.2020}, supernova ejecta \citep{Rho.2022}, and massive star-forming regions by ALMA \citep{Cortes.2024,Sanhueza.2025} and protoplanetary disks\citep{Stephens:2014gv}. In such environments, dust temperatures can reach $T_d \gtrsim 100$–$200$ K, and the magnetic response of grains may differ substantially from that in the cold ISM. Whether dust polarization reliably traces magnetic fields and how the alignment efficiency varies under these conditions depends sensitively on the temperature- and frequency-dependent magnetic susceptibility of grains \citep{LazHoang.2007,HoangLaz.2016,Hoangetal.2022}.

\citet{Hoang.2025} recently generalized RAT theory to provide a unified description of grain alignment and rotational disruption across diverse astrophysical environments, highlighting the central role of radiation field and magnetic relaxation in the efficiency of alignment and disruption. In this paper, we develop a physically motivated model for the effective magnetic susceptibility of dust grains containing SPIs with a power-law size distribution. We quantify how the effective susceptibility depends on dust temperature and grain angular frequency, identify the characteristic inclusion size that dominates the magnetic response, and derive analytic expressions for the resulting frequency spectrum. We then explore the implications of this effective superparamagnetism for magnetic grain alignment and magnetic dipole emission across diverse astrophysical environments.

The paper is organized as follows. Section \ref{sec:magrev} reviews the magnetic properties of paramagnetic and superparamagnetic material and discuss constraints on the maximum size of SPIs. In Section \ref{sec:effective_suscept}, we present our model for the effective susceptibility for grains with size-distributed SPIs. Section \ref{sec:alignment} discusses the effects of effective susceptibility for grain alignment. Discussion and a summary of our main findings are shown in Section \ref{sec:discuss} and \ref{sec:summary}.

\section{Magnetic Properties of Astrophysical Dust}\label{sec:magrev}
In this section, we provide a brief overview of the fundamental theory concerning the magnetic properties of astrophysical dust for reference. We begin with paramagnetic and superparamagnetic dust, focusing on the case of uniform-sized SPIs, before moving on to discuss SPI size distribution in the following section.

\subsection{Paramagnetic Dust}
When iron atoms are diffusely distributed within the matrix of a dust grain, they create a standard paramagnetic material (PM). The zero-frequency susceptibility of this paramagnetic dust at temperature $T_{d}$ is described by Curie's law.
\bea
\chi_{\rm PM}(0)&=&\frac{n_{p}\mu^{2}}{3k_{\B}T_{\d}},\label{eq:curielaw_PM}
\ena
where $n_{p}$ is the number density of paramagnetic (Fe) atoms inside the dust grain, and $\mu$ is the effective magnetic moment per iron atom $\mu$ given by
\bea
\mu^{2}\equiv p^{2}\mu_{\B}^{2} =g_{e}^{2}\mu_{B}^{2}\left[{j(j+1)}\right],\label{eq:mu}
\ena
with $j$ being the angular momentum quantum number of electrons in the outer partially filled shell, and $p\approx 5.5$ is taken for silicate (see \citealt{Draine.1996}).

Plugging the typical numerical values into Equation (\ref{eq:curielaw_PM}), we obtain
\bea
\chi_{\rm PM}(0)\simeq 0.03f_{p}\hat{n}_{23}\left(\frac{p}{5.5}\right)^{2}\left(\frac{20\K}{T_{d}}\right),\label{eq:chi_PM}
\ena
where $\hat{n}_{23}=n/10^{23}\cm^{-3}$ is the normalized density, $f_{p}=n_{p}/n$ is the fraction of Fe atoms in the matrix of silicate grains with $f_{p}=1/7$ for silicate of structure MgFeSiO$_{4}$.


\subsection{Superparamagnetic Dust}
Iron particles (including metallic (Fe) and iron oxide particles) at low temperatures tend to be ferromagnetic in which the exchange interaction between electron spins (magnetic dipoles) enables them to be spontaneously aligned in an easy axis (e.g, long axis) for which the crystalline anisotropy energy is minimized. This results in spontaneous magnetization $M_{s}$ with $4\pi M_{s}=22000, 6400, 4780~G$ for Fe, Fe$_{3}$O$_{4}$ and $\gamma$Fe$_{2}$O$_{3}$ (see, e.g., \citealt{DraineLaz:1999} (hereafter DL99) for more details).  When the temperature exceeds some threshold, so-called blocking temperature, $T_{\rm block}$, the particle becomes superparamagnetic because thermal fluctuations of the lattice exceed the energy barrier (i.e., magnetostatic energy) to randomize the individual magnetic moments, resulting in a zero net intrinsic magnetic moment. 

The incorporation of iron nanoparticles (hereafter iron clusters) can substantially enhance the magnetic susceptibility of a dust grain. When the grain temperature of $T_d > T_{\rm block}$, the magnetic moments of single-domain clusters within the grain fluctuate rapidly, and the composite grain behaves as a superparamagnetic material (SPM). Let $N_{\rm cl}$ denote the number of structural formula units of the magnetic material (i.e., Fe, Fe$_{2}$O$_{3}$, or Fe$_{3}$O$_{4}$) contained in a superparamagnetic cluster. For a metallic Fe cluster of spherical shape and radius $r$ with mass density $\rho=7.86\g\cm^{-3}$, $N_{\rm cl}=(4\pi/3) r^{3}\rho/m_{\rm Fe}\simeq 355r_{-7}^{3}$ with $r_{-7}=r/10^{-7}\cm$.



\subsection{Magnetic Susceptibility for Grains with Uniform SPIs}

For magnetic materials, the response of the magnetization $M(t)$ to an oscillating magnetic field is described by dynamical equations. To model the magnetic response of the ferromagnetic material, DL99 employed the Drude model with scalar susceptibility, in which the magnetic response to the oscillating field $H(t)$ is described by
\bea
\ddot{M}(t)=\omega_{0}^{2}\left[\chi(0){H(t)}- {M(t)} \right]- \frac{\dot{M}(t)}{\tau_{0}},\label{eq:ddMdt}
\ena
where $\chi(0)$ is the magnetic susceptibility at zero frequency, $\omega_{0}$ is the resonance angular velocity (or frequency) which is defined as the precession frequency of an electron in the internal field of $B_{\rm int}=4\pi M_{s}/3$, i.e., $\omega_{0}=(eB_{\rm int}/m_{e}c)$ and $\tau_{0}$ is the characteristic magnetic relaxation (damping) time. 

The complex magnetic susceptibiltiy is described by $\chi(\omega)=\chi_{1}(\omega) + i\chi_{2}(\omega)$ where $\chi_{1}$ and $\chi_{2}$ are the real and imaginary part. The solution of Equation (\ref{eq:ddMdt}) reads (see {DL99}):
\bea
\chi_{1}(\omega)=\frac{\chi(0)\left[1-(\omega/\omega_{0})^{2}\right]}{[1-(\omega/\omega_{0})^{2}]^{2}+(\omega \tau)^{2}},\\
\chi_{2}(\omega)=\frac{\chi(0)\omega\tau}{[1-(\omega/\omega_{0})^{2}]^{2}+(\omega \tau)^{2}},\label{eq:chi2}
\ena
which depends crucially on the product $\omega_{0}\tau$. For $\omega_{0}\tau<2$, $\chi(\omega)$ has the resonance near $\omega_{0}$, but the system is damped significantly for $\omega_{0}\tau>2$. The critically damped state occurs at $\omega_{0}\tau=2$ or $\omega_{0}\tau_{0}=2$. 
 
Assuming the critically-damped condition with $\omega_{0}\tau=2$ (DL99), the above equations become
\bea
\chi_{1}^{cd}(\omega)&=&\frac{\chi(0) \omega \tau\left[1-(\omega\tau/2)^{2}\right]}{[1+(\omega \tau/2)^{2}]^{2}}~ \label{eq:chi1_cd},\\
\chi_{2}^{cd}(\omega)&=&\frac{\chi(0) \omega \tau}{[1+(\omega \tau/2)^{2}]^{2}} ~~\label{eq:chi2_cd},
\ena
where $\tau=(\omega_{0}^{2}\tau_{0})^{-1}$ is the effective relaxation time, and the denominators indicate that the magnetic susceptibility is significantly suppressed when the frequency of the applied field becomes larger than the response rate of the system (i.e., $\omega > \omega_{0}=\tau^{-1}$).

For our case of grains rotating at angular frequency $\omega$, although the grains are subject to a static interstellar magnetic field, the magnetic field appears rotating within the grain frame of reference at frequency $\omega$. Therefore, Equations (\ref{eq:chi1_cd}-\ref{eq:chi2_cd}) can be used to describe the magnetic susceptibility as a function of the grain angular frequency.

Let consider first the case of uniform SPIs having equal size of $N_{\rm cl}$. In thermal equilibrium, the effective magnetic moment of the ensemble of SPIs can be described by the Langevin function with argument $m H/k_{B}T_{d}$, where $m= N_{\rm cl}\mu_{0}$ is the total magnetic moment of the cluster and $\mu_{0}$ is the magnetic moment of basic magnetic unit (e.g., Fe atom for metallic iron), and $H$ is the applied magnetic field (\citealt{JonesSpitzer.1967}). Let $\mu_{0}=p\mu_{B}$ where $p$ is the number of Bohr magneton per basic magnetic unit. The exact values $p$ depends on the cluster size and it varies from $p\sim 3$ for small metallic clusters $N_{\rm cl}<100$ and saturates the the bulk metallic value of $p\sim 2.2$ for $N_{\rm cl}>500$ at $T_{d}=120\K$ \citep{Billas:1994p6378}. As the temperature increases, $p$ decreases to $p\approx 0.5$ for $T_{d}\sim 500-900\K$.\footnote{Lower values of $p=1.25$ and $1.18$ for Fe$_{2}$O$_{3}$ and Fe$_{3}$O$_{4}$ (see Table 1 in \cite{Yang.2021}).}

For the case of uniform iron clusters of size $N_{\rm cl}$, the composite grain exhibits superparamagnetic behavior, which has the magnetic susceptibility given by
\bea
\chi_{\rm SPM}(0,N_{\rm cl})=\frac{n_{\rm cl}m^{2}}{3k_{B}T_{d}},\label{eq:chisp_rest}
\ena
where $n_{\rm cl}$ is the volume density of iron clusters (iron clusters per unit volume). 

Let $\phi_{\rm SPI}$ be the volume filling factor of SPIs within the grain. Throughout this paper, for numerical estimates, we assume the typical value of $\phi_{\rm SPI}=0.03$, i.e., about 10$\%$ of cosmic iron in the form of iron clusters \citep{Bradley:1994p6379,HoangLaz.2016}. Plugging $m$ and $n_{\rm cl}=\mathcal{N}/V_{\rm grain}$ with the total number of clusters {$\mathcal{N}=\phi_{\rm SPI}V_{\rm grain}/(N_{\rm cl}V_{\rm Fe})$}$\approx 3.5\times 10^{8}\phi_{\rm SPI}N_{\rm cl}^{-1}a_{-5}^{3}$ into Equation (\ref{eq:chisp_rest}), one obtains:
\bea
\chi_{\rm SPM}(0,N_{\rm cl})&& = \frac{\phi_{\rm SPI}N_{\rm cl}}{3V_{\rm Fe}}\frac{p^{2}\mu_{B}^{2}}{k_{B}T_{d}},\nonumber\\
&&\approx 0.011\phi_{\rm SPI}N_{\rm cl}\hat{p}^{2}\hat{T}_{d}^{-1},\label{eq:chi_Ncl_zero}
\ena
where $a_{-5}=a/10^{-5}\cm$, $\hat{p}=p/3$, and $\hat{T}_{d}\equiv T_{d}/15\K$.

SPIs of size $N_{\rm cl}$ undergo remagnetization due to thermal excitations, and the relaxation rate (aka. N\'eel relaxation rate) is given by (see \citealt{DraineLaz.1999}, \citealt{Morrish:2001vp})
\bea
\tau^{-1}_{N}=\nu_{0}\exp\left(-KV_{\rm cl}/k_{B}T_{d}\right)= \nu_0 \exp\left(-N_{\rm cl}T_{\rm act}/T_{d}\right), \label{eq:tau_sp}
\ena
with $\nu_{0}$ the attempt frequency, $KV_{\rm cl}$ the energy required for remagnetization of iron clusters of volume $V_{\rm cl}$, and $T_{\rm act}$ is the activation temperature of superparamagnetism defined by $T_{\rm act}=KV_{\rm cl}/(N_{\rm cl}k_{B})$. For metallic Fe clusters with $K=1.35 \times 10^{5}{\rm erg}\cm^{-3}$ and $V_{\rm cl}=N_{\rm cl}V_{\rm Fe}=1.2\times 10^{-23}N_{\rm cl}\boldsymbol{\cm^{3}}$ where $V_{\rm Fe}=1.18 \times 10^{-23}\cm^{3}$ is the atomic Fe volume, one obtains $\boldsymbol{T_{\rm act}}=KV/(k_{B}V_{\rm Fe})=0.011\K$. Similarly, Fe$_{2}$O$_{3}$ (maghemite, $K=6.1\times 10^{5}{\rm erg}\cm^{-3}, V_{Fe_{2}O_{3}}=7.2\times 10^{-23}\cm^{3}$), $T_{\rm act}=KV_{\rm cl}/(k_{B}V_{Fe_{2}O_{3}})=0.32\K$. For Fe$_{3}$O$_{4}$ (magnetite, $K=1.9\times 10^{5}{\rm erg}\cm^{-3}, V_{Fe_{3}O_{4}}=7.4\times 10^{-23}\cm^{3}$), $T_{\rm act}=KV_{\rm cl}/(k_{B}V_{Fe_{3}O_{4}})=0.1\K$. In the following, our numerical results are shown for SPIs of metallic irons, but the results can be adapted for maghemite and magnetite.

Metallic iron is the most favorable material for superparamagnetism, so our numerical results for metallic Fe SPIs should be regarded as upper limits. In astrophysical environments, iron in SPIs may be largely oxidized into magnetite or maghemite, which have larger activation temperatures and lower spontaneous magnetizations. Because the critical blocking size (Eq. \ref{eq:Ncr}) and the resonance size (Eq. \ref{eq:Npeak}) scale as $T_{d}/T_{\rm act}$, they are smaller by factors of $\approx 9$ and $\approx 30$ (in numbers of formula units) for magnetite and maghemite, respectively. Writing the cluster moment as $m=M_{s}V_{\rm cl}$ in Equation (\ref{eq:chisp_rest}) and using $T_{\rm act}=KV_{\rm cl}/(N_{\rm cl}k_{B})$, the susceptibility of SPIs at the critical blocking size becomes
\bea
\chi_{\rm SPM}(0,N_{\rm cri})=\frac{\phi_{\rm SPI}M_{s}^{2}}{3K}\ln(\nu_{0}\tau_{\rm ran}),\label{eq:chimax_material}
\ena
which is independent of $T_{d}$ and depends on the material only through $M_{s}^{2}/K$; the same holds for $\chi_{\rm SPM}(0,N_{\rm res})$, with $\ln(\nu_{0}/\omega)$ in place of $\ln(\nu_{0}\tau_{\rm ran})$. Since $F_{\rm eff}$ and $G_{\rm eff}$ depend only weakly on the material, the effective susceptibility and dissipation strength of magnetite (maghemite) SPIs with the same $\phi_{\rm SPI}$ are lower than those of metallic Fe SPIs by a factor of $\approx 17$ ($\approx 100$) for the values of $M_{s}$ and $K$ given above, or $\approx 30$ ($\approx 170$) relative to the moment $p=3$ adopted here for metallic Fe clusters, i.e., $M_{s}=p\mu_{B}/V_{\rm Fe}$. 


\subsection{Critical blocking size of SPIs determined by magnetic relaxation and gas randomization}

Thermal fluctuations enable the magnetic moments of iron clusters to undergo N\'eel relaxation, allowing composite grains to exhibit SPM behavior. For this to influence grain alignment, the relaxation timescale, $\tau_{N}$, must be shorter than the randomization timescale due to gas collisions $\tau_{\rm gas}$ (see Eq.~\ref{eq:tgas}). If $\tau_{N}>\tau_{\rm gas}$, grains are randomized before SPIs can relax, corresponding to the regime in which SPIs are effectively “blocked”. Because $\tau_{N}$ depends on SPI size, we define the critical size above which SPIs are blocked (the critical blocking size), $N_{\rm cri}$, by setting $\tau_{N}=\tau_{\rm ran}$:
\bea
N_{\rm cri}=\left(\frac{T_{d}}{T_{\rm act}}\right)\ln \left(\nu_{0}\tau_{\rm gas}\right)\simeq 909.1\left(\frac{T_{d}}{10\K}\right)\ln\left(\nu_{0}\tau_{\rm ran}\right).\label{eq:Ncr}
\ena

Using the randomization time $\tau_{\gas}\sim (8.26\times 10^{6}/n_{\H}) \yr$ (Eq. \ref{eq:tgas}) and
$\nu_{0}\sim 10^{9}\s^{-1}$, one has $\ln(\nu_{0}\tau_{\rm gas})\approx 44.7, 21.7$ for $n_{\H}=10^{4}\cm^{-3}$ and $n_{\H}=10^{14}\cm^{-3}$. The critical blocking size of SPIs is then $N_{\rm cri}=4.06\times 10^{4}(T_{d}/10\K)$ and $1.97\times 10^{4}(T_{d}/10\K)$, which decreases by a factor of 2 when the density increases by 10 orders of magnitude. Therefore, dust temperature is the main parameter controling the critical blocking size, $N_{\rm cri}$.

For $N_{\rm cl}>N_{\rm cri}$, SPIs are blocked and the SPM of these clusters are zero. Therefore, the maximum susceptibility of superparamagnetic grains is given by
\bea
\chi_{\rm SPM}^{\rm max}(0)=\chi_{\rm SPM}(0,N_{\rm cri}),\label{eq:chimax_Ncr}
\ena
calculated by Equation (\ref{eq:chi_Ncl_zero}) in which $N_{\rm cl}= N_{\rm cri}$, corresponding to the case in which all iron clusters embedded in the grain have the same size.

\section{Effective magnetic susceptibility of Superparamagnetic Grains}\label{sec:effective_suscept}
\subsection{Size distribution of SPIs}
Here we consider a realistic case in which SPIs follow a distribution function $f(N)$ with $N$ the number of iron atoms per iron cluster.\footnote{Here, $N\equiv N_{\rm cl}$, but the subscript is omitted for convenience.} The number density of SPIs per unit physical volume of the host grain with size between $N, N+dN$ is given by the power-law distribution
\bea
dn = f(N)dN=CN^{-q}dN,\label{eq:dn_dN}
\ena 
where $f_{N}=CN^{-q}$ is the size distribution with the slope $q>0$, and $C$ is the normalization factor determined by the iron abundance in the form of iron clusters (i.e., volume filling factor $\phi_{\rm SPI}$). The SPI size distribution has the lower limit $N_{\rm min}$ to upper limit $N_{\rm max}$. The lower limit is $N_{\rm min}\sim 20$ \cite{Billas:1994p6378}, and the upper limit is determined by the upper limit for single-domain ferromagnetic material of $N_{\rm max}\approx 5\times 10^{5}$ \citep{Kneller:1963p6410}. For this paper, we {adopt a fiducial slope} of $q=11/6$ (see Appendix \ref{apdx:Cconst}), derived from the assumption that {the radii of} SPIs follow the standard size distribution of interstellar dust \citep{Mathis:1977p3072}. {Because this assumption is not constrained by observations or laboratory data, we also show results for $q=13/6$ and $15/6$ (Figure \ref{fig:chi2_zero_Ncl_avg_Td}) and for log-normal size distributions (Appendix \ref{apdx:lognormal_spi}).}

For the power-law SPI size distribution, the normalization constant $C$ is given by (see Appendix \ref{apdx:Cconst}), 
\bea
C= \frac{(2-q)\phi_{\rm SPI}}{V_{\rm Fe}N_{\rm max}^{2-q}}\frac{1}{\left[1-(N_{\rm min}/N_{\rm max})^{2-q}\right]},\label{eq:Cnorm}
\ena
where $V_{\rm Fe}$ is the atomic Fe volume; both $dn$ and $C$ have units of $\cm^{-3}$.

\subsection{Effective magnetic susceptibility at zero-frequency}
The magnetic susceptibility of SPM grains induced by iron clusters with size between $N, N+dN$ is given by
\bea
d\chi(0,N)=\frac{N^{2}p^{2}\mu_{B}^{2}dn}{3k_{B}T_{d}}=\frac{N^{2}p^{2}\mu_{B}^{2}f(N)dN}{3k_{B}T_{d}}.\label{eq:dchi0}
\ena

We define the {\it effective superparamagnetic susceptibility} as the susceptibility obtained by averaging the single-particle superparamagnetic response over the size distribution of SPIs, which becomes
\bea
\chi_{\rm eff}(0)&&=\int_{N_{\rm min}}^{N_{\rm max}}d\chi(0,N)=\int_{N_{\rm min}}^{N_{\rm max}}\frac{p^{2}\mu_{B}^{2}N^{2}f(N)dN}{3k_{B}T_{d}},\nonumber\\
&&=\int_{N_{\rm min}}^{N_{\rm max}}\frac{Cp^{2}\mu_{B}^{2}N^{2-q}dN}{3k_{B}T_{d}},\label{eq:chi_SPM_avg_zero1}
\ena
which can be rewritten as
\bea
\chi_{\rm eff}(0) &&=\frac{C p^{2}\mu_{B}^{2}}{3k_{B}T_{d}}\int_{N_{\rm min}}^{N_{\rm cri}}N^{2-q}dN\nonumber\\
&&=\frac{C p^{2}\mu_{B}^{2}}{3(3-q)k_{B}T_{d}}N_{\rm cri}^{3-q}\left[1-(N_{\rm min}/N_{\rm cri})^{3-q}\right].\label{eq:chi_SPM_avg_zero2}
\ena

Equation (\ref{eq:chi_SPM_avg_zero2}) is similar to the finding in \cite{Yang.2021} who derived it using a different approach.

Using the normalization $C$ from Equation (\ref{eq:Cnorm}), one obtains
\bea
 \chi_{\rm eff}(0) &&= \chi_{\rm SPM}(0,N_{\rm cri})\times F_{\rm eff},\nonumber\\
 &&=\chi_{\rm max}(0)\times F_{\rm eff},\label{eq:chi_avg_zero}
\ena
where $\chi_{\rm max}(0)\equiv \chi_{\rm SPM}(0,N_{\rm cri})$ is the maximum superparamagnetic susceptibility estimated at $N_{\rm cl}=N_{\rm cri}$ given by Equation (\ref{eq:chi_SPM_avg_zero1}), and the reduction factor $F_{\rm eff}$ given by
\bea
F_{\rm eff}&&= \frac{2-q}{3-q}\left(\frac{N_{\rm cri}}{N_{\rm max}}\right)^{2-q}\left(\frac{1-(N_{\rm min}/N_{\rm cri})^{3-q}}{1-(N_{\rm min}/N_{\rm max})^{2-q}} \right),\nonumber \\
&&=\frac{2-q}{3-q}\left(\frac{T_{d}}{T_{\rm act}}\right)^{2-q}\left(\frac{\ln(\nu_{0}\tau_{\rm ran})}{N_{\rm max}}\right)^{2-q}\nonumber\\
&& \times \left(\frac{1-(N_{\rm min}/N_{\rm cri})^{3-q}}{1-(N_{\rm min}/N_{\rm max})^{2-q}} \right).\label{eq:F_SPM}
\ena

Since $N_{\rm min}\ll N_{\rm max}$ and $N_{\rm min}\ll N_{\rm cri}$, the final bracket term goes to unity. For $q<2$, the reduction factor $F_{\rm eff}$ increases with $N_{\rm cri}$. Because $N_{\rm cri}\sim T_{d}$, the effective susceptibility varies as $\chi_{\rm eff}(0) \propto T_{d}^{3-q}/T_{d}\sim T_{d}^{2-q}$, which increases with $T_{d}$ for the standard slope of $q=11/6$. This is opposite to the steep decrease as $1/T_{d}$ for the single-size SPI case given by Curie law (\ref{eq:chi_Ncl_zero}).

Equations (\ref{eq:chi_SPM_avg_zero2})--(\ref{eq:F_SPM}) assume $N_{\rm cri}\leq N_{\rm max}$, i.e., $T_{d}\leq T_{\rm sat}\equiv N_{\rm max}T_{\rm act}/\ln(\nu_{0}\tau_{\rm ran})\approx 110$--$120\K$ for $\ln(\nu_{0}\tau_{\rm ran})\approx 45$--$51$. At higher temperatures, $N_{\rm cri}>N_{\rm max}$ and all SPIs are thermally active, so the upper limit of the integral in Equation (\ref{eq:chi_SPM_avg_zero2}) is $N_{\rm max}$, and $\chi_{\rm eff}(0)=\chi_{\rm SPM}(0,N_{\rm max})F_{\rm eff}(N_{\rm max})$ with
\bea
F_{\rm eff}(N_{\rm max})=\frac{2-q}{3-q}\left(\frac{1-(N_{\rm min}/N_{\rm max})^{3-q}}{1-(N_{\rm min}/N_{\rm max})^{2-q}} \right),\label{eq:F_SPM_sat}
\ena
which gives $F_{\rm eff}(N_{\rm max})\approx 0.18$ for $q=11/6$. In this saturated regime, the effective susceptibility follows Curie's law, $\chi_{\rm eff}(0)\propto 1/T_{d}$.

The maximum SPI size is also uncertain. We adopt $N_{\rm max}=5\times10^{5}$, the upper limit for single-domain particles \citep{Kneller:1963p6410}, but SPIs need not reach this size in all environments, e.g., in cold or iron-poor regions. The dependence on $N_{\rm max}$ is weak only while $N_{\rm cri}<N_{\rm max}$: then $\chi_{\rm eff}(0)\propto N_{\rm max}^{-(2-q)}$ (Eq. \ref{eq:F_SPM}), so that a ten times smaller $N_{\rm max}$ increases $\chi_{\rm eff}(0)$ by a factor of $\approx 1.6$ for $q=11/6$, because the same amount of iron is then contained in smaller, active SPIs. However, $N_{\rm max}$ also determines $T_{\rm sat}\propto N_{\rm max}$. For $N_{\rm max}=5\times10^{4}$, $T_{\rm sat}\approx 11$--$12\K$, and $\chi_{\rm eff}(0)$ at $T_{d}=100$--$300\K$ is lower by a factor of $\approx 8$ than for $N_{\rm max}=5\times10^{5}$. For rotating grains, the rapid decrease of $\chi_{\rm eff}(\omega)$ once $N_{\rm res}$ reaches $N_{\rm max}$ (see below Eq. \ref{eq:G_SPM}) then sets in at $T_{d}\approx 48\K$ and $80\K$, instead of $\approx 480\K$ and $800\K$, for $\omega=10^{4}$ and $10^{6}\s^{-1}$. Thus, a smaller $N_{\rm max}$ mainly lowers the temperatures above which the effective susceptibility decreases.

\begin{figure*}
	\includegraphics[width=0.5\textwidth]{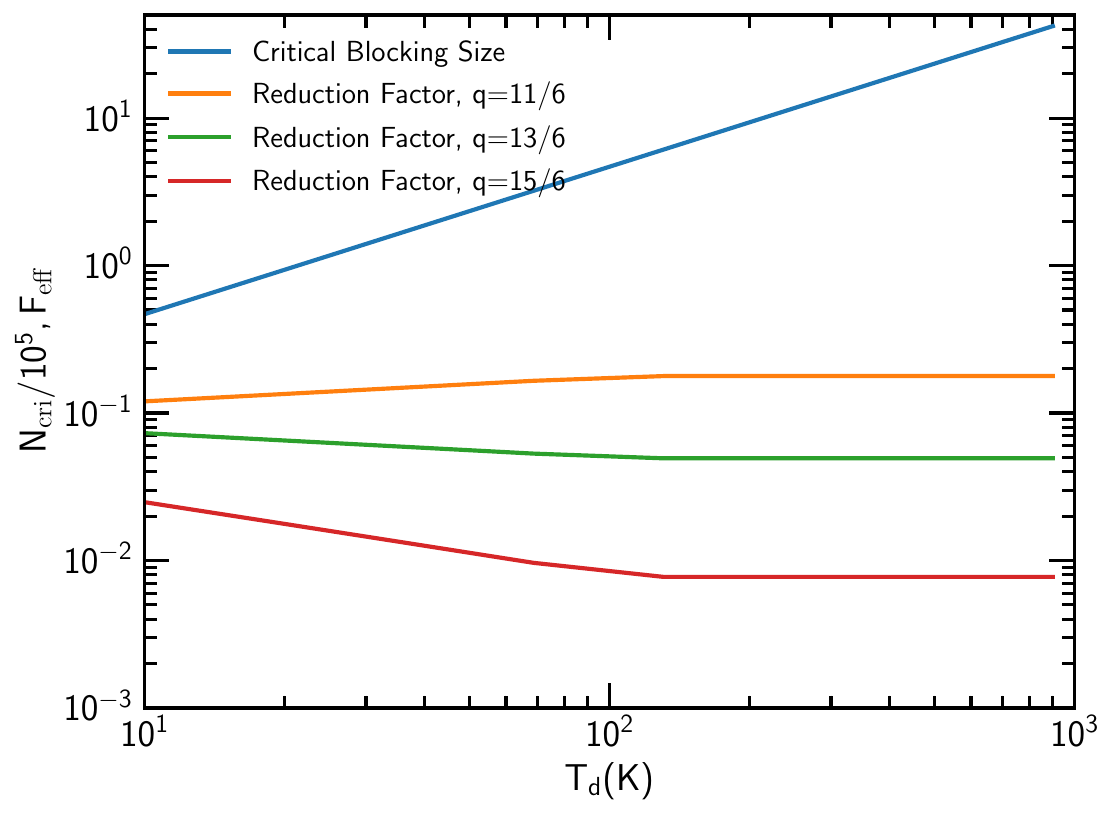}
    \includegraphics[width=0.5\textwidth]{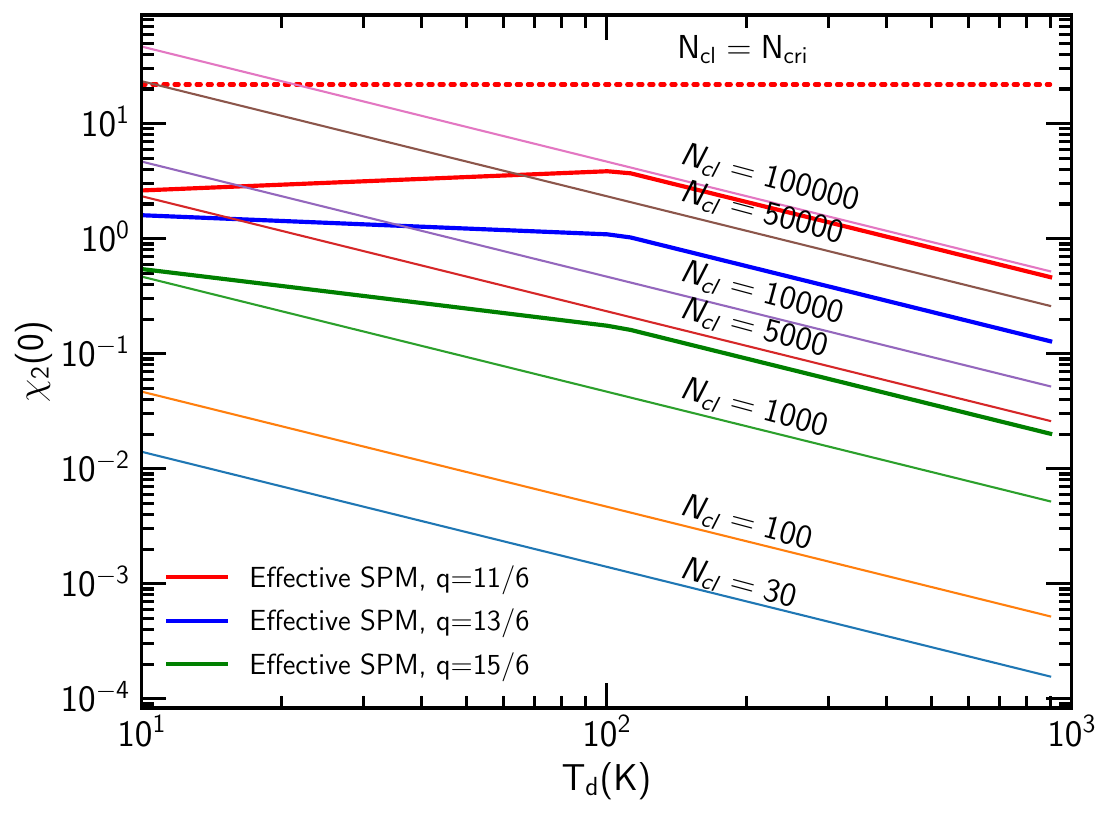}
    \caption{Left panel: The variation of the critical blocking size of SPIs and the effective reduction factor with $T_{d}$ for slopes of $q=11/6, 13/6, {15/6}$. Right panel: The variation of the zero-frequency susceptibility with the dust temperature for the single-size SPIs compared to the effective susceptibility and the maximum susceptibility at the blocking size $N_{\rm cri}$. {The maximum susceptibility is independent of $T_{d}$, and the effective susceptibility varies slowly with $T_{d}$ (increasing for $q=11/6$) until $N_{\rm cri}$ reaches $N_{\rm max}$ at $T_{d}\approx 110\K$, beyond which it decreases as $1/T_{d}$, in contrast to the decrease as $1/T_{d}$ at all temperatures for the single-size SPIs. The randomization time is $\tau_{\rm ran}=\tau_{\rm gas}$ for $a=1\mum$, $n_{\H}=20\cm^{-3}$, and $T_{\rm gas}=100\K$ ($\ln(\nu_{0}\tau_{\rm ran})=51.4$).}}
    \label{fig:chi2_zero_Ncl_avg_Td}
\end{figure*}

Figure \ref{fig:chi2_zero_Ncl_avg_Td} (left panel) shows the rapid increase of the critical blocking SPI size with $T_{d}$, but the reduction factor $F_{\rm eff}$ {varies slowly with $T_{d}$ (increasing for $q=11/6$ and decreasing for $q=13/6$ and $15/6$), spanning $\sim 0.01$--$0.2$ for three slopes of the SPI size distribution, and becomes constant for $T_{d}>T_{\rm sat}$ where $N_{\rm cri}>N_{\rm max}$}. It implies that larger SPIs are activated by increasing the dust temperature.

Figure \ref{fig:chi2_zero_Ncl_avg_Td} (right panel) shows the zero-frequency susceptibility as a function of dust temperature for different SPI sizes ($\chi_{2}(0,N_{\rm cl})$), the maximum susceptibility at the critical size ($\chi_{\rm SPM}(0, N_{\rm cri})$), and the effective susceptibilty obtained by integrating over the SPI size distribution ($\chi_{\rm eff}(0)$) for the three different slopes $q$. The single-size susceptibility decreases with temperature as $1/T_{d}$ described by Curie's law. In contrast, {the maximum susceptibility $\chi_{\rm SPM}(0,N_{\rm cri})$ is independent of $T_{d}$ because $N_{\rm cri}\propto T_{d}$, and} the effective susceptibility varies only slightly with $T_{d}$ {for $T_{d}<T_{\rm sat}$} due to the increase in the critical blocking size induced by stronger thermal fluctuations. For the typical slope of $q=11/6$, the effective susceptibility even slightly increases with $T_{d}$, in contrast to Curie's law. {Above $T_{\rm sat}\approx 110\K$, all SPIs are active, and $\chi_{\rm eff}(0)$ decreases as $1/T_{d}$ (Eq. \ref{eq:F_SPM_sat}).}

\subsection{Rotating Grains and Resonance Size of SPIs}
For a grain rotating with the angular velocity $\omega$, the frequency-dependence susceptibility (imaginary part) of SPIs of size $N_{\rm cl}$ in the critically damped condition is given by
\bea
\chi_{2}^{cd}(\omega,N_{\rm cl}) = \frac{\chi(0) \omega \tau_{N}}{[1+(\omega \tau_{N}/2)^{2}]^{2}},\label{eq:chi2_cd_Ncl}
\ena
where is the magnetic relaxation timescale for SPIs of size $N_{\rm cl}$, as given by Equation (\ref{eq:tau_sp}). Because only the imaginary part of the magnetic susceptibility is relevant for magnetic grain alignment and magnetic dipole emission, in the following, we present the results for the imaginary part only. The theory can be straightforwardly applied to the real part.

The magnetic susceptibility $\chi_{2}^{cd}(\omega,N_{\rm cl})$ achieves its maximum at the rotation rate of $\omega=\tau_{N}^{-1}$, defined as resonance rotation rate
\bea
\omega_{\rm res}=\nu_{0}\exp\left(-\frac{T_{\rm act}N_{\rm cl}}{T_{d}}\right),\label{eq:omega_res}
\ena
which corresponds to $\omega_{\rm res}=1.7\times 10^{4}\s^{-1}$ for $N_{\rm cl}=10^{4}$, and $\omega_{\rm res}=8.9\times 10^{8}\s^{-1}$ for $N_{\rm cl}=100$, assuming the same temperature of $T_{d}=10\K$.

For a given grain rotation rate and a SPI size, one can define the resonance temperature as
\bea
T_{\rm res}=\frac{T_{\rm act}N_{\rm cl}}{\ln\left(\nu_{0}/\omega\right)},\label{eq:Tres}
\ena
which increases with increasing $N_{\rm cl}$ and rotation rate $\omega$. For the large SPI of $N_{\rm cl}=10^{5}$, the resonance temperature is $T_{\rm res}=95.5\K, 119.4\K$, and $159.2\K$ for $\omega=10^{4}, 10^{5}$ and $10^{6}\s^{-1}$, respectively. Therefore, large SPIs have a maximum susceptibility when the grain temperature reaches $T_{\rm d}\sim 100-150\K$.

For an ensemble of SPIs of different sizes embedded within the grain, we can define the size of SPIs that undergoes resonance as the SPI {\it resonance size}:
\bea
N_{\rm res} = \frac{T_{d}}{T_{\rm act}}\ln\left(\frac{\nu_{0}}{\omega}\right)\simeq 909.1\left(\frac{T_{d}}{10\K}\right)\ln\left(\frac{10^{9}\s^{-1}}{\omega}\right),\label{eq:Npeak}
\ena
which increases with $T_{d}$ but decreases with $\omega$. SPIs of resonance size have dominant contribution to the susceptibility, whereas SPIs larger than $N_{\rm res}$ have thermal fluctuations slower than the grain rotation and have a minor contribution to the magnetic susceptibility.

The magnetic dissipation strength depends on the imaginary part of susceptibility as $K_{\rm SPM}(\omega,N_{\rm cl})=\chi_{2}(\omega,N_{\rm cl})/\omega$, which becomes:
\bea
K_{\rm SPM}(\omega,N_{\rm cl}) &&= \frac{\chi_{\rm SPM}(0)\tau_{N}}{[1+(\omega \tau_{N}/2)^{2}]^{2}}\nonumber\\
&& \simeq  \boldsymbol{1.6\times 10^{-10}}\hat{p}^{2}N_{\rm cl}\phi_{\rm SPI}\times \frac{k_{\rm SPM}(\omega)}{T_{d}},~~~\label{eq:kappa_spm}
\ena
where the last term denotes the dependence on $T_{d}$ and the angular frequency with
\bea
k_{\rm SPM}(\omega,N_{\rm cl})= \exp\left(\frac{N_{\rm cl}T_{\rm act}}{T_{d}}\right)\left[1+\left(\frac{\omega\tau_{N}}{2}\right)^{2}\right]^{-2},\label{eq:gsp}
\ena
which increases with $N_{\rm cl}$ but decreases with increasing $T_{d}$ and $\omega$

For rotating grains at $\omega<\tau_{N}^{-1}$, the second term in the denominator is negligible and
\bea
\frac{k_{\rm SPM}}{T_{d}}\sim \left(\frac{\exp\left(N_{\rm cl}T_{\rm act}/T_{d}\right)}{T_{d}}\right),\label{eq:kSPM_Td}
\ena
which decreases with temperature more rapid than the zero-frequency susceptibility given by Curie's law. Therefore, strong thermal fluctuations (N\'eel magnetic relaxation, i.e., smaller $\tau_{N}$) act to significantly decrease the SPM susceptibility.



\subsection{Effective magnetic susceptibility for the SPI size distribution}
Here we calculate the effective susceptibility for the case of SPI size distribution. Following the similar procedure as above, the imaginary part of the magnetic susceptibility induced by iron clusters of size $N,N+dN$ is given by
\bea
d\chi_{2}(\omega,N)=d\chi(0,N)\frac{\omega \tau_{N}}{[1+(\omega \tau_{N}/2)^{2}]^{2}},\label{eq:dchi_omega}
\ena
where $\tau_{N}$ is the relaxation time of iron clusters $N$ from Equation (\ref{eq:tau_sp}).

The effective superparamagnetic susceptibility is given by
\bea
\chi_{\rm eff}(\omega) &&=\int_{N_{\rm min}}^{N_{\rm max }}d\chi_{2}(\omega,N) \nonumber\\
&&=\int_{N_{\rm min}}^{N_{\rm max }} \frac{p^{2}\mu_{B}^{2}}{3kT} \frac{\omega \tau_{N}}{[1+(\omega \tau_{N}/2)^{2}]^{2}}N^{2}f(N)dN, \label{eq:chi_SPM_omega_avg}
\ena
which yields

\bea
\chi_{\rm eff}(\omega) = \frac{Cp^{2}\mu_{B}^{2}}{3k_{B}T_{d}}\int_{N_{\rm min}}^{N_{\rm max }} \frac{\omega \tau_{N}}{[1+(\omega \tau_{N}/2)^{2}]^{2}}N^{2-q}dN.\label{eq:chi_SPM_omega_avg_int}
\ena

For $N\ll N_{\rm res}$ ($\omega\tau_{N}\ll 1$), the integrand in Equation (\ref{eq:chi_SPM_omega_avg_int}) is proportional to $\omega\tau_{N}\propto e^{NT_{\rm act}/T_{d}}$ and grows exponentially with $N$, whereas for $N\gg N_{\rm res}$ it is suppressed as $(\omega\tau_{N})^{-3}$. The integral is therefore dominated by SPIs in a narrow range of width $\Delta N\sim T_{d}/T_{\rm act}\ll N_{\rm res}$ around $N_{\rm res}$. Changing the variable to $y=\omega\tau_{N}=\exp(x-x_{\rm res})$, with $x=NT_{\rm act}/T_{d}$ and $x_{\rm res}=N_{\rm res}T_{\rm act}/T_{d}=\ln(\nu_{0}/\omega)$, Equation (\ref{eq:chi_SPM_omega_avg_int}) becomes
\bea
\chi_{\rm eff}(\omega) &&= \frac{Cp^{2}\mu_{B}^{2}}{3k_{B}T_{d}}\left(\frac{T_{d}}{T_{\rm act}}\right)^{3-q}\nonumber\\
&&\times\int_{y_{1}}^{y_{2}} \frac{x(y)^{2-q}dy}{[1+(y/2)^{2}]^{2}},\label{eq:chi_eff_y}
\ena
where $x(y)=x_{\rm res}+\ln y$, $y_{1}=\exp(x_{1}-x_{\rm res})$ and $y_{2}=\exp(x_{2}-x_{\rm res})$, with $x_{1}=N_{\rm min}T_{\rm act}/T_{d}$ and $x_{2}=N_{\rm max}T_{\rm act}/T_{d}$. Because the kernel is concentrated at $y\sim 1$, where $|\ln y|\lesssim 1\ll x_{\rm res}$, the factor $x^{2-q}$ can be replaced by $x_{\rm res}^{2-q}$. For $N_{\rm min}\ll N_{\rm res}\lesssim N_{\rm max}$, i.e., $y_{1}\ll 1\ll y_{2}$, the integral can also be extended to $[0,\infty)$, where $\int_{0}^{\infty}dy/[1+(y/2)^{2}]^{2}=\pi/2$. Using $(T_{d}/T_{\rm act})^{3-q}x_{\rm res}^{2-q}=(T_{d}/T_{\rm act})N_{\rm res}^{2-q}$, one obtains
\bea
\chi_{\rm eff}(\omega) \simeq \frac{\pi}{2}\frac{Cp^{2}\mu_{B}^{2}}{3k_{B}T_{\rm act}}N_{\rm res}^{2-q}.\label{eq:chi_eff_closed}
\ena

Using the normalization $C$ from Equation (\ref{eq:Cnorm}), this can be rewritten as
\bea
\chi_{\rm eff}(\omega) &&=\chi_{\rm SPM}(0,N_{\rm res})\times G_{\rm eff}(\omega),
\label{eq:chi_avg_omega_final}
\ena
where $\chi_{\rm SPM}(0,N_{\rm res})$ is the zero-frequency susceptibility of single-size SPIs at the resonance size $N_{\rm res}$ given by Equation (\ref{eq:chi_Ncl_zero}), and $G_{\rm eff}$ is the effective reduction factor given by
\bea
G_{\rm eff}(\omega)&&\simeq\frac{\pi}{2}\left(\frac{2-q}{1-(N_{\rm min}/N_{\rm max})^{2-q}}\right)\nonumber\\
&&\times\left(\frac{N_{\rm res}}{N_{\rm max}}\right)^{2-q}\frac{1}{\ln(\nu_{0}/\omega)}.\label{eq:G_SPM}
\ena

Equation (\ref{eq:G_SPM}) agrees with the numerical integration of Equation (\ref{eq:chi_SPM_omega_avg_int}) to within $\sim 1\%$ for $N_{\rm min}\ll N_{\rm res}\lesssim N_{\rm max}$. The numerical integration also satisfies the exact sum rule $\int_{0}^{\infty}\chi_{\rm eff}(\omega)d\omega/\omega=(\pi/2)\chi_{\rm eff}(0)$ (without blocking), which follows from Equation (\ref{eq:chi2_cd_Ncl}) for each SPI size. For the standard slope of $q=11/6$, $G_{\rm eff}\approx 0.016$--$0.043$ for $10^{4}\leq\omega\leq 10^{6}\s^{-1}$ and $15\K\leq T_{d}\leq 500\K$ (where $N_{\rm res}<N_{\rm max}$), i.e., the effective susceptibility is a few percent of the single-size susceptibility at $N_{\rm res}$. Because $\chi_{\rm SPM}(0,N_{\rm res})\propto N_{\rm res}/T_{d}=\ln(\nu_{0}/\omega)/T_{\rm act}$, Equations (\ref{eq:chi_avg_omega_final}) and (\ref{eq:G_SPM}) yield $\chi_{\rm eff}(\omega)\propto N_{\rm res}^{2-q}\propto [T_{d}\ln(\nu_{0}/\omega)]^{2-q}$. Therefore, the effective magnetic susceptibility is mostly determined by $N_{\rm res}$ instead of $N_{\rm cri}$ as in the case of grains at rest (Eq. \ref{eq:chi_SPM_avg_zero2}). It depends only weakly on the frequency, as $[\ln(\nu_{0}/\omega)]^{2-q}$, and on the temperature, as $T_{d}^{2-q}$, which increases with $T_{d}$ for the standard slope of $q=11/6$. When $N_{\rm res}$ reaches $N_{\rm max}$, i.e., for $T_{d}\gtrsim N_{\rm max}T_{\rm act}/\ln(\nu_{0}/\omega)\approx 480\K$ and $800\K$ for $\omega=10^{4}$ and $10^{6}\s^{-1}$, all SPIs relax faster than the grain rotation ($\omega\tau_{N}<1$), and $\chi_{\rm eff}(\omega)$ decreases rapidly with increasing $T_{d}$.

The effective superparamagnetic dissipation strength is given by
\bea
K_{\rm eff}(\omega) &&=\frac{\chi_{\rm eff}(\omega)}{\omega}= \chi_{\rm SPM}(0,N_{\rm res})\times \frac{G_{\rm eff}(\omega)}{\omega}\nonumber\\
&&\simeq\frac{\pi}{2}\left(\frac{2-q}{1-(N_{\rm min}/N_{\rm max})^{2-q}}\right)\nonumber\\
&&\times\left(\frac{N_{\rm res}}{N_{\rm max}}\right)^{2-q}\frac{\chi_{\rm SPM}(0,N_{\rm res})}{\omega\ln(\nu_{0}/\omega)},\label{eq:Kappa_avg_omega_final}
\ena
where Equation (\ref{eq:G_SPM}) has been used.

\begin{table*}[]
    \centering
        \caption{List of symbols and meaning}
    \begin{tabular}{ll}
    \hline
Notation     & Meaning \\

    \hline\hline
$T_{d}$ & Dust grain temperature \\
$\omega$ & Grain rotation rate or angular frequency\\
$\phi_{\rm SPI}$ & Volume filling factor of iron clusters inside the grain\\
$N_{\rm cl}$    & Number of iron atoms per cluster \\
$N_{\rm min},N_{\rm max}$ & Minimum and maximum number of iron atoms per cluster \\
$N_{\rm cri}$ & Critical blocking size of SPIs determined by N\'eel relaxation and grain randomization \\
$\tau_{N}$ & Characteristic N\'eel relaxation timescale of SPIs by thermal fluctuations\\
$\chi_{\rm SPM}(0,N_{\rm cl})$ & Zero-frequency magnetic susceptibility of the single-size SPIs \\
$\chi_{\rm SPM}(\omega, N_{\rm cl})$ & Frequency-dependence superparamagnetic susceptibility for the single-size SPIs \\
$\chi_{\rm eff}(\omega)$ & Effective superparamagnetic susceptibility for the SPI size distribution \\
$K_{\rm eff}(\omega)=\chi_{\rm eff}(\omega)/\omega$ & Effective superparmagnetic dissipation strength for the SPI size distribution\\
$N_{\rm res}=T_{d}/T_{\rm act}\ln(\nu_{0}/\omega)$ & Resonance size of SPIs\\
$T_{\rm res}=\frac{T_{\rm act}N_{\rm cl}}{\ln\left(\nu_{0}/\omega\right)}$ & Resonance grain temperature\\
$\omega_{\rm res}=\nu_{0}\exp\left(-\frac{T_{\rm act}N_{\rm cl}}{T_{d}}\right)$ & Resonance grain angular frequency\\
\hline\hline
    \end{tabular}
    \label{tab:tab_symbol}
\end{table*}



\begin{figure}
	\includegraphics[width=0.5\textwidth]{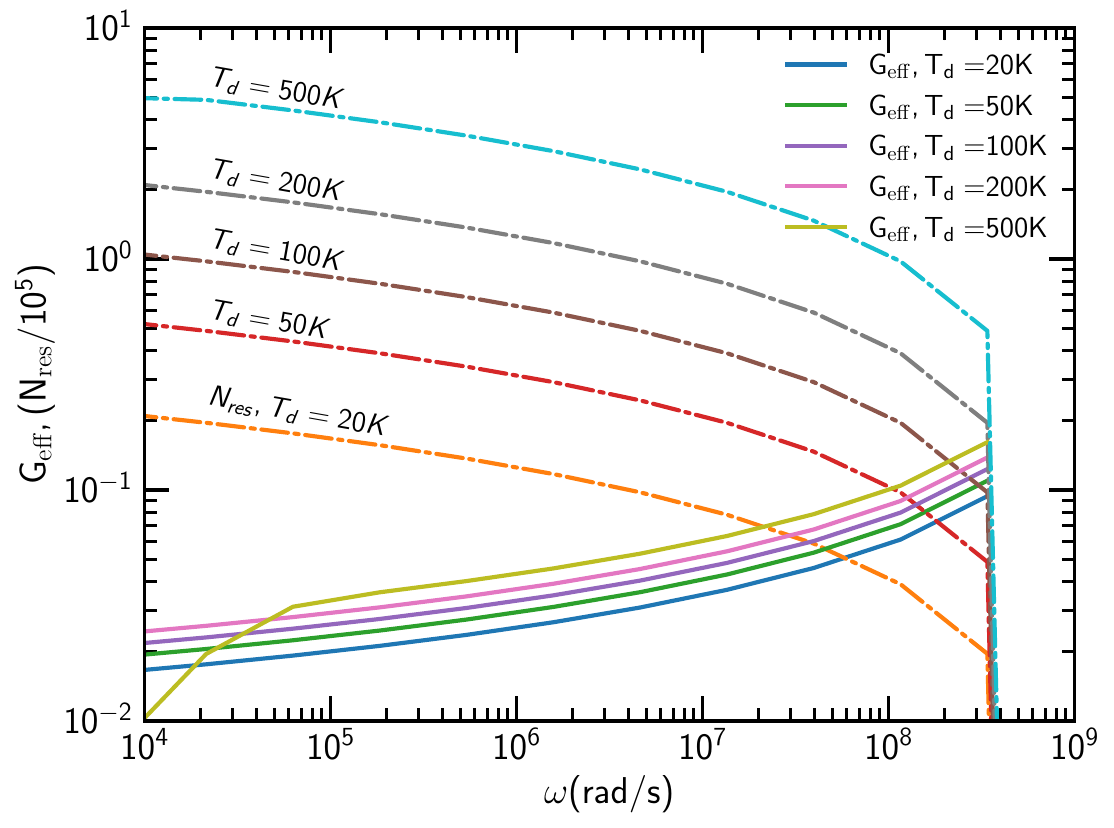}
    \caption{Variation of the resonance SPI size ($N_{\rm res}$) and reduction factor $G_{\rm eff}$ as a function of the grain rotation rate for the different dust temperature. For realistic rotation rates of $\omega<10^{6}\s^{-1}$, $N_{\rm res}$ increases from $10^{4}$ to the maximum SPI size of $N_{\rm max}=5\times 10^{5}$ for $T_{d}$ increases from $15\K$ to $500\K$.}
    \label{fig:GSPM_omega_Td}
\end{figure}

Figure~\ref{fig:GSPM_omega_Td} shows the dependence of the resonance size of SPIs, $N_{\rm res}$, and the reduction factor $G_{\rm eff}$ on the grain rotation rate for different dust temperatures. For a fixed rotation rate, $N_{\rm res}$ increases with dust temperature, ranging from $\sim 10^{4}$ at $T_d = 15\,\mathrm{K}$ to the maximum SPI size $N_{\rm max} = 5 \times 10^{5}$ at $T_d = 900\,\mathrm{K}$. The reduction factor $G_{\rm eff}$ increases with both angular frequency and dust temperature, reflecting the enhanced contribution of thermally activated inclusions. {For $10^{4}\leq\omega\leq 10^{6}\s^{-1}$ and $T_d \leq 500\,\mathrm{K}$, $G_{\rm eff}\approx 0.016$–$0.043$ (Eq. \ref{eq:G_SPM})}, indicating that the effective susceptibility remains a {small fraction} of the maximal single-size value.

\begin{figure*}
\centering
	\begin{overpic}[width=0.45\textwidth]{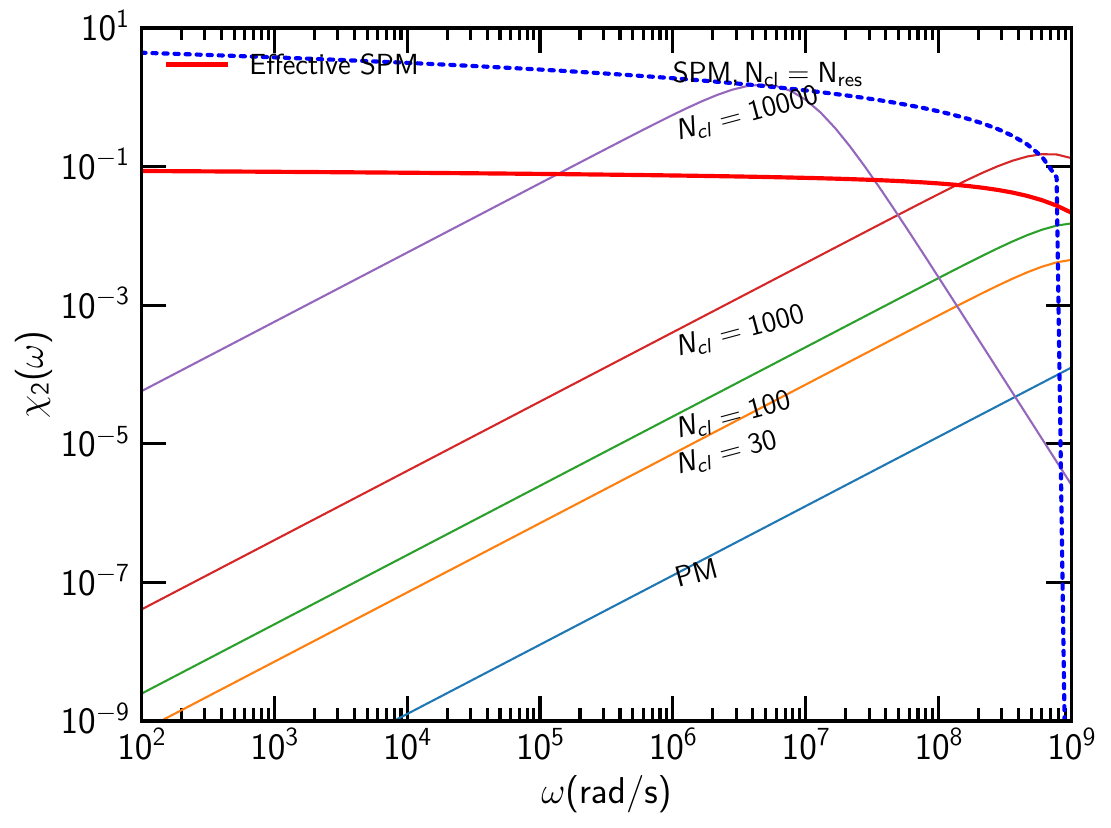}
    \put(65,15){\small \textbf{(a)} $T_{d}=20$K}
    \end{overpic}
\begin{overpic}[width=0.45\textwidth]{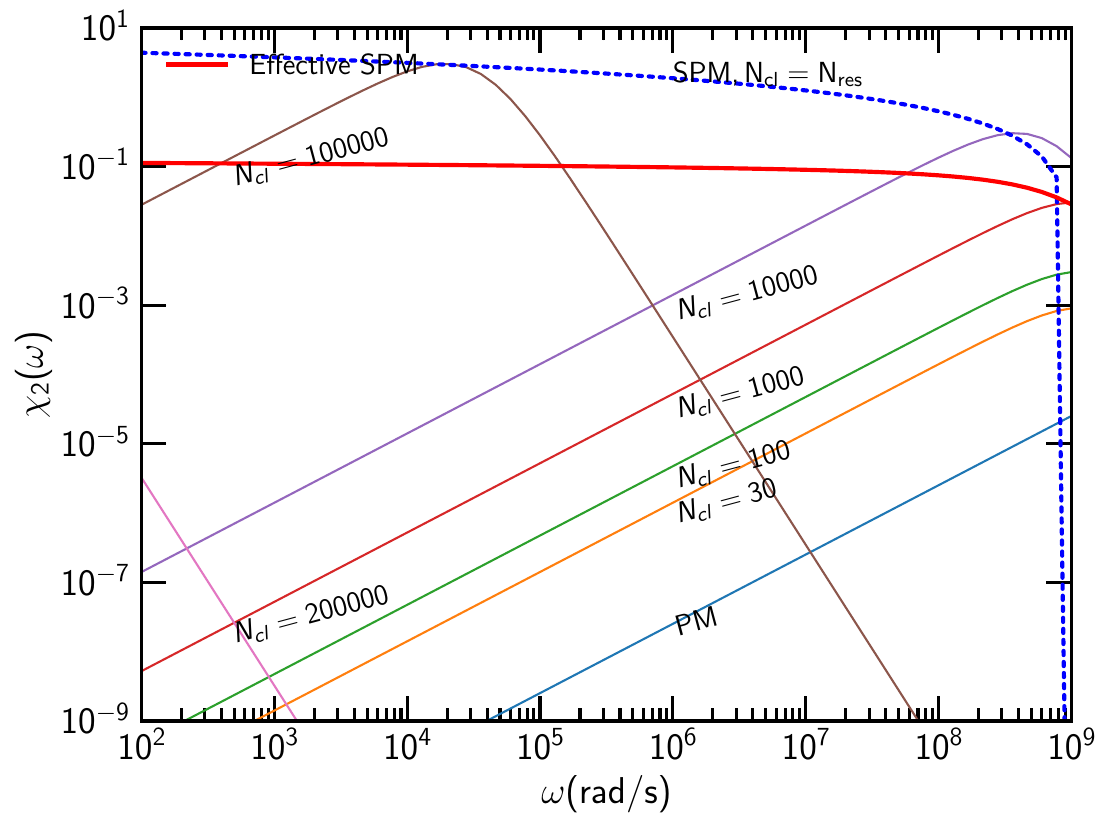}
     \put(65,15){\small \textbf{(b)} $T_{d}=100$K}
      \end{overpic}
\begin{overpic}[width=0.45\textwidth]{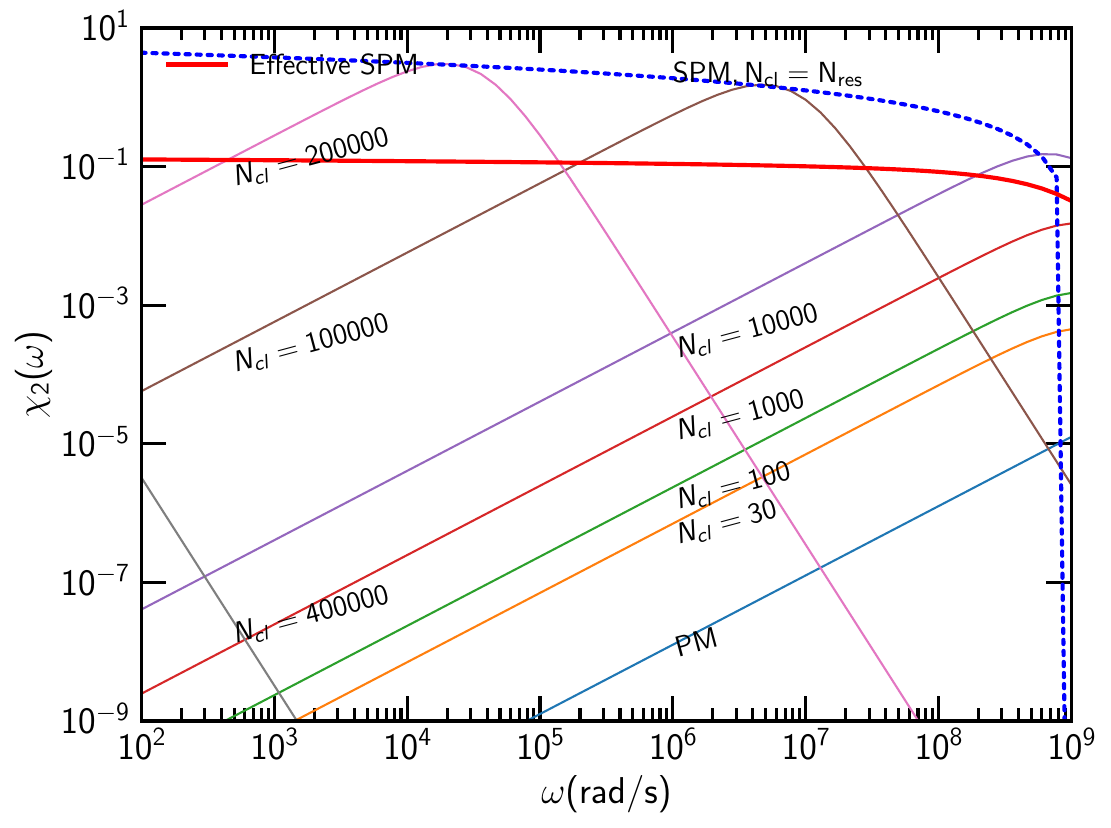}
     \put(65,15){\small \textbf{(c)} $T_{d}=200$K}
      \end{overpic}
\begin{overpic}[width=0.45\textwidth]{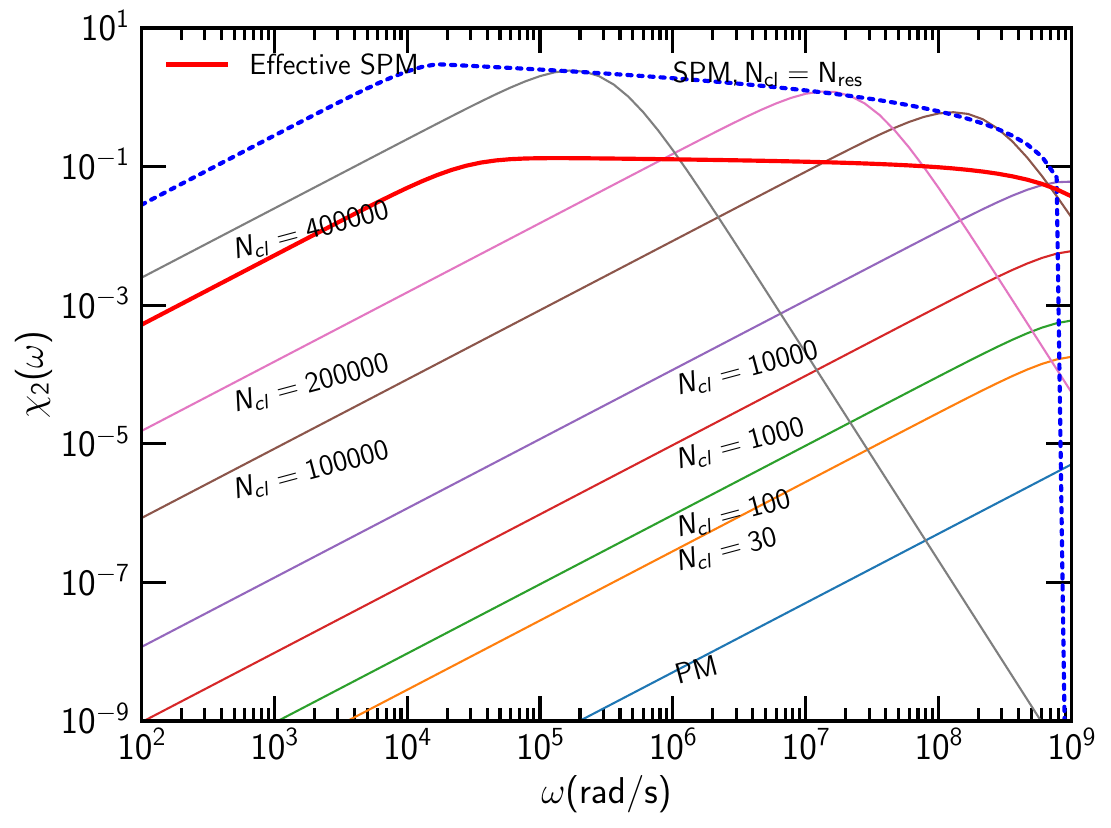}
     \put(65,15){\small \textbf{(d)} $T_{d}=500$K}
      \end{overpic}
	\caption{The imaginary part of the magnetic susceptibility, $\chi_{2}(\omega, N_{\rm cl})$, for superparamagnetic grains with uniformly sized SPIs as a function of grain rotation rate for different values of $N_{\rm cl}$ and dust temperatures {$T_d = 20, 100, 200, 500\K$} (thin solid lines). For comparison, {the susceptibility of single-size SPIs at the resonance size, $\chi_{2}(\omega,N_{\rm res})$} (thin dotted line), and the effective susceptibility (thick solid line) are also plotted. As the rotation rate decreases from the characteristic $\nu_0=10^{9}\s^{-1}$, the peak of $\chi_{2}(N_{\rm cl}, \omega)$ shifts toward larger $N_{\rm cl}$ due to the resonance effect, while both the maximum and effective susceptibilities slightly increase.}
    \label{fig:chi2_omega_Td}
\end{figure*}

Figure~\ref{fig:chi2_omega_Td} illustrates the frequency dependence of the superparamagnetic susceptibility, $\chi_{\rm SPM}(\omega,N_{\rm cl})$, for uniform-sized SPIs at different dust temperatures. For comparison, we also show the effective susceptibility, $\chi_{\rm eff}(\omega)$, obtained from the size-distribution model, as well as the paramagnetic (PM) susceptibility. For uniform-sized clusters, $\chi_{\rm SPM}(\omega,N_{\rm cl})$ has the maximum at the resonance frequency $\omega_{\rm res}$ that shifts to lower values as the cluster size $N_{\rm cl}$ increases. This behavior follows from the resonance condition $\omega_{\rm res} \sim \tau_{N}^{-1} \propto \exp(-N_{\rm cl} T_{\rm act}/T_d)$, indicating the exponential sensitivity of the N\'eel relaxation time to cluster size and temperature. In contrast, the effective susceptibility $\chi_{\rm eff}(\omega)$ exhibits a flat frequency dependence and remains nearly constant over a broad range of $\omega$, until the angular frequency approaches the characteristic attempt frequency $\nu_0$. This behavior arises from the superposition of resonant contributions from SPIs of different sizes.

At higher temperatures, such as $T_d = 500\K$, larger SPIs become thermally activated. In this regime, the effective susceptibility increasingly resembles that of the largest active SPIs and shows a steeper decline at low rotation rates, particularly for $\omega \lesssim 10^{4}\,\mathrm{rad\,s^{-1}}$ (bottom-right panel).

\begin{figure*}
	\includegraphics[width=0.5\textwidth]{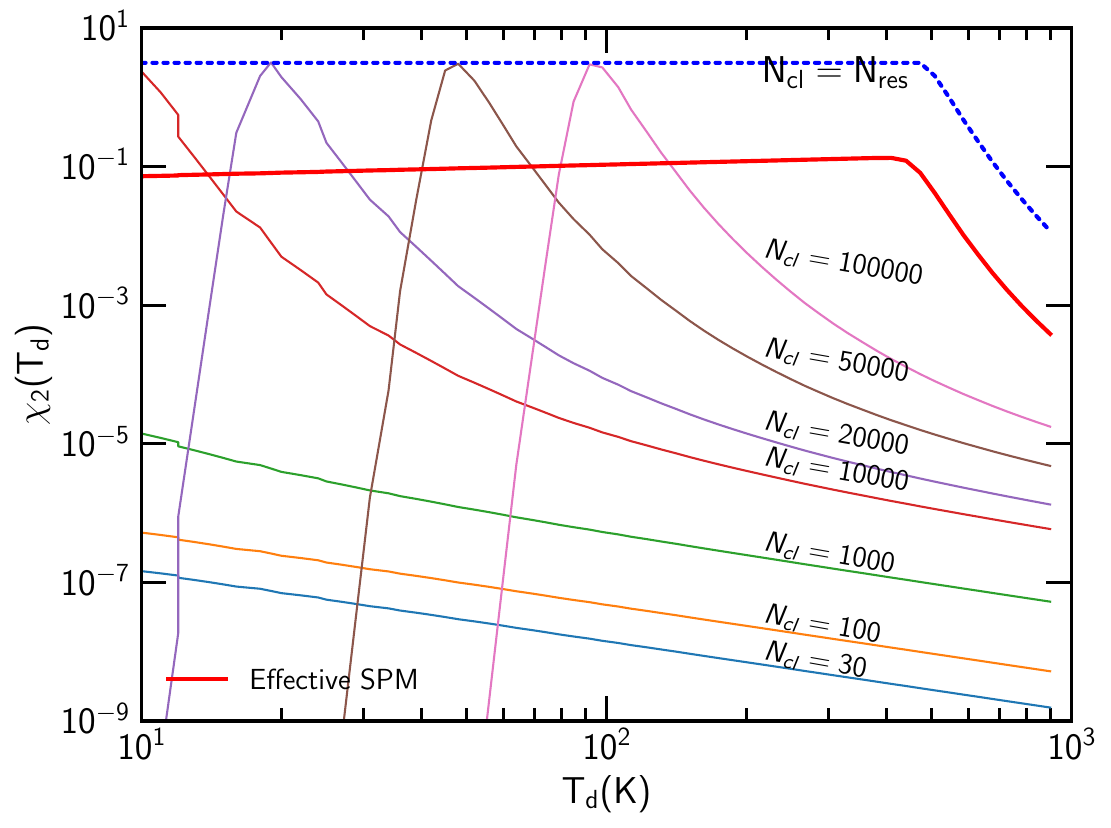}
	\includegraphics[width=0.5\textwidth]{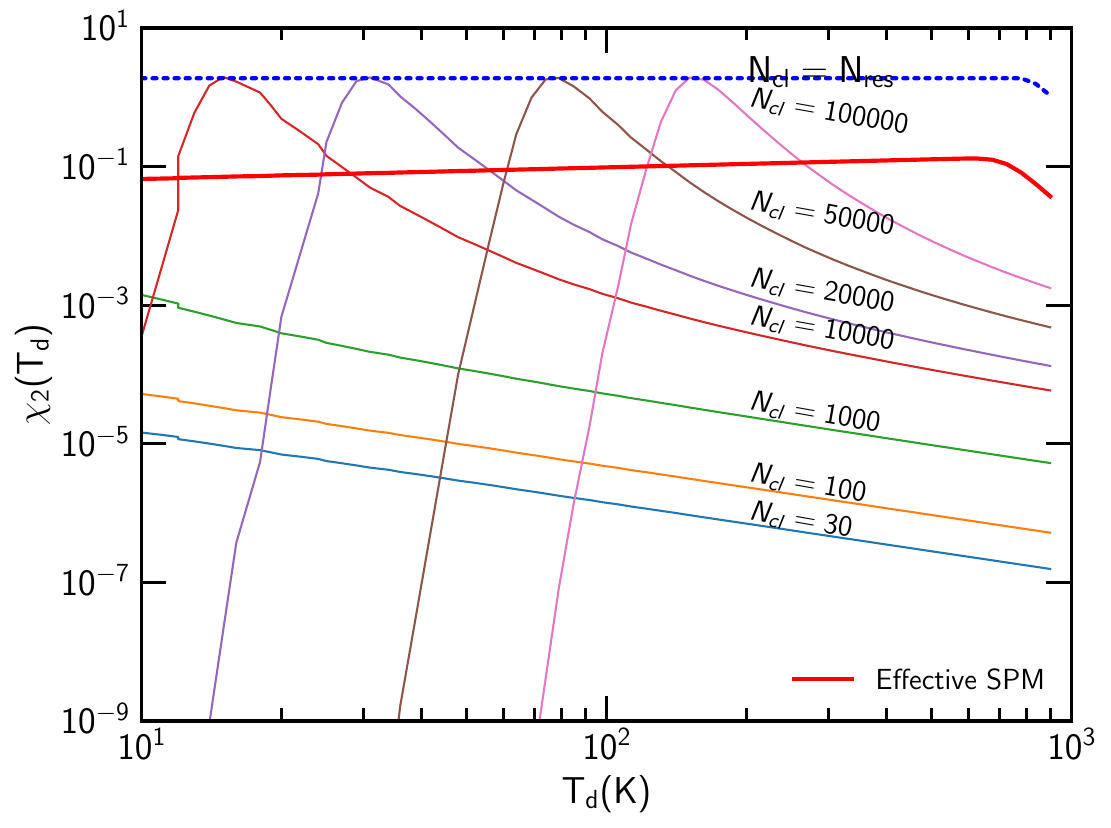}
	\caption{Variation of $\chi_{2}(\omega, N_{\rm cl})$ and effective susceptibility as functions of $T_{d}$ for different iron clusters, assuming the grain rotation rate of $\omega=10^{4}\s^{-1}$ (left panel) and $10^{6}\s^{-1}$ (right panel). As the temperature increases, the peak of $\chi_{2}(N_{\rm cl})$ shifts toward larger $N_{\rm cl}$ due to stronger thermal fluctuations. The effective susceptibility slightly increases with $T_{d}$, in contrast to the resonant behavior at $T_{d}=T_{\rm res}$ in the case of the uniform-sized SPIs{, until $N_{\rm res}$ reaches $N_{\rm max}$ ($T_{d}\approx 480\K$ and $800\K$ for $\omega=10^{4}$ and $10^{6}\s^{-1}$), beyond which it decreases rapidly.}}
    \label{fig:chi2_avg_Td_Ncl_omega}
\end{figure*}

Figure \ref{fig:chi2_avg_Td_Ncl_omega} shows the variation of $\chi_{\rm SPM}(\omega,N_{\rm cl})$ with $T_{d}$ for the different $N_{\rm cl}$ and at two grain angular frequencies. For the grains with single SPI size, the susceptibility has its maximum at $T_{d}=T_{\rm res}$ (Eq. \ref{eq:Tres}) and decreases rapidly beyond its peak as described by $\exp(N_{\rm cl}T_{\rm act}/T_{d})/T_{d}$ (see Equation \ref{eq:kSPM_Td}). However, the effective susceptibility slowly increases with $T_{d}${, as $T_{d}^{2-q}$,} resulting from the activation of larger SPIs at higher $T_{d}${, until $N_{\rm res}$ reaches $N_{\rm max}$.}

\begin{figure*}
    \centering
	\begin{overpic}[width=0.45\textwidth]{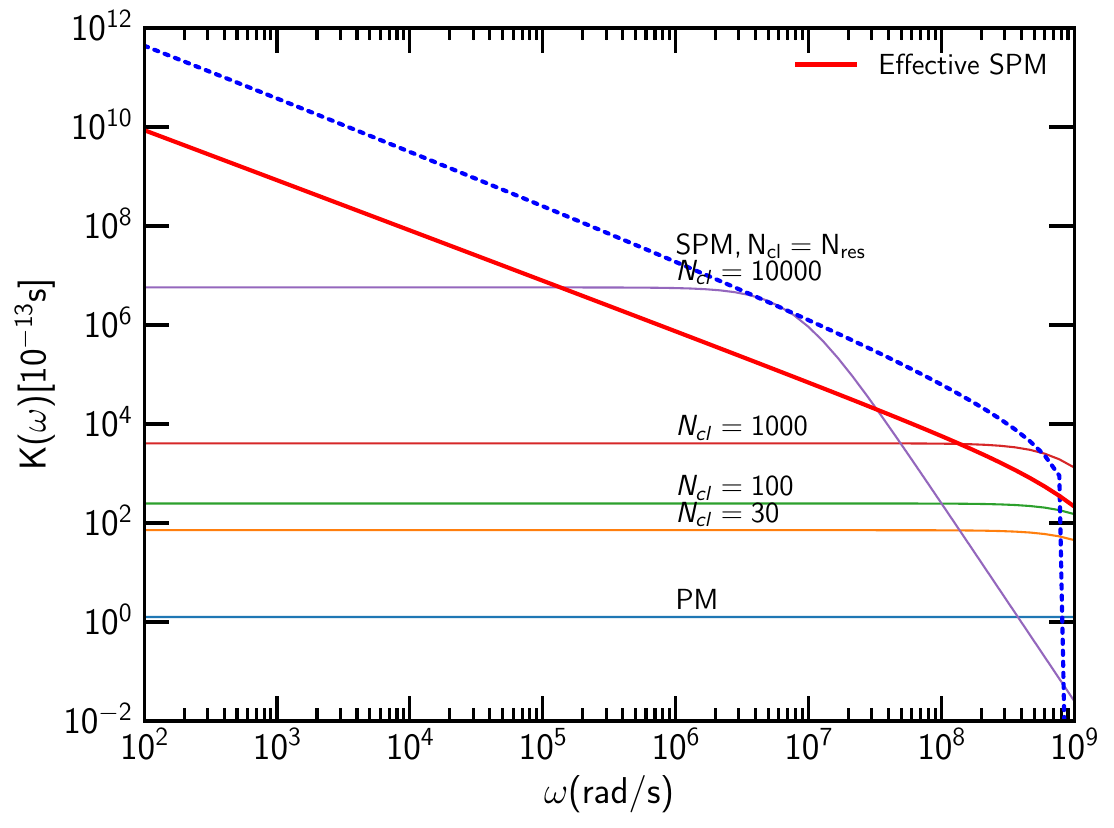}
    \put(70,60){\small \textbf{(a)} $T_{d}=20$K}
    \end{overpic}
\begin{overpic}[width=0.45\textwidth]{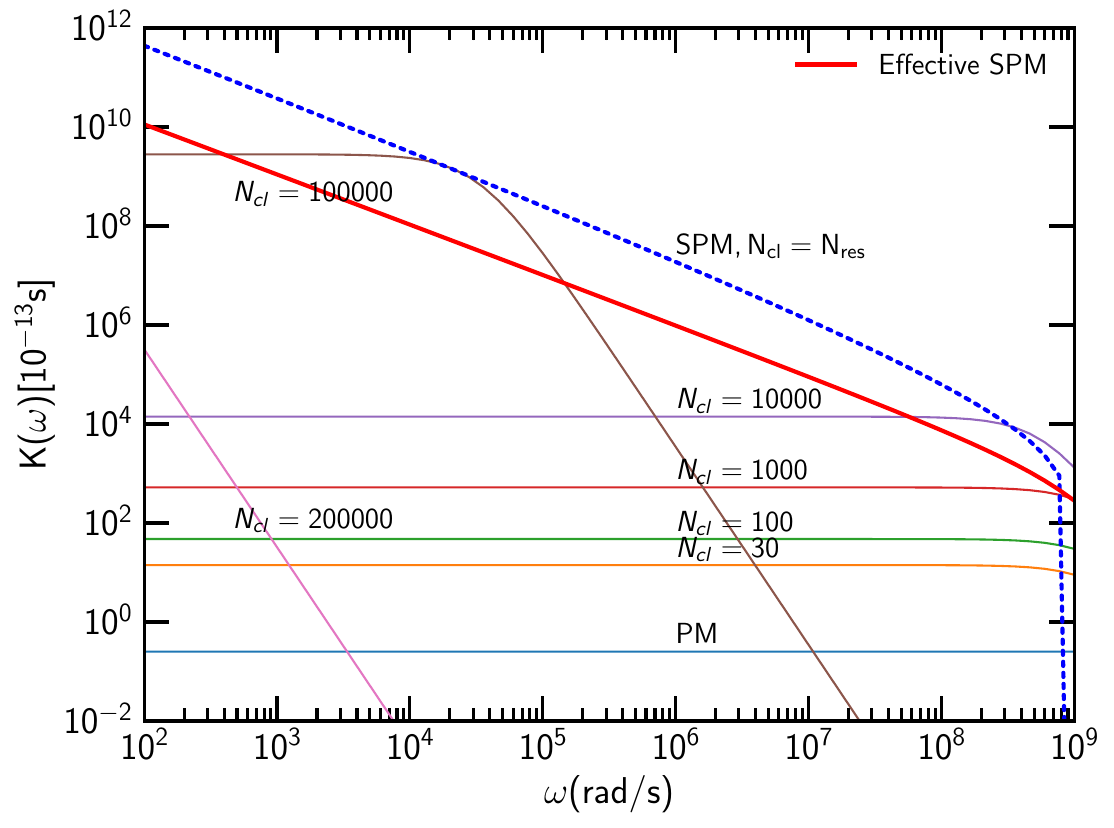}
 \put(70,60){\small \textbf{(b)} $T_{d}=100$K}
      \end{overpic}
\begin{overpic}[width=0.45\textwidth]{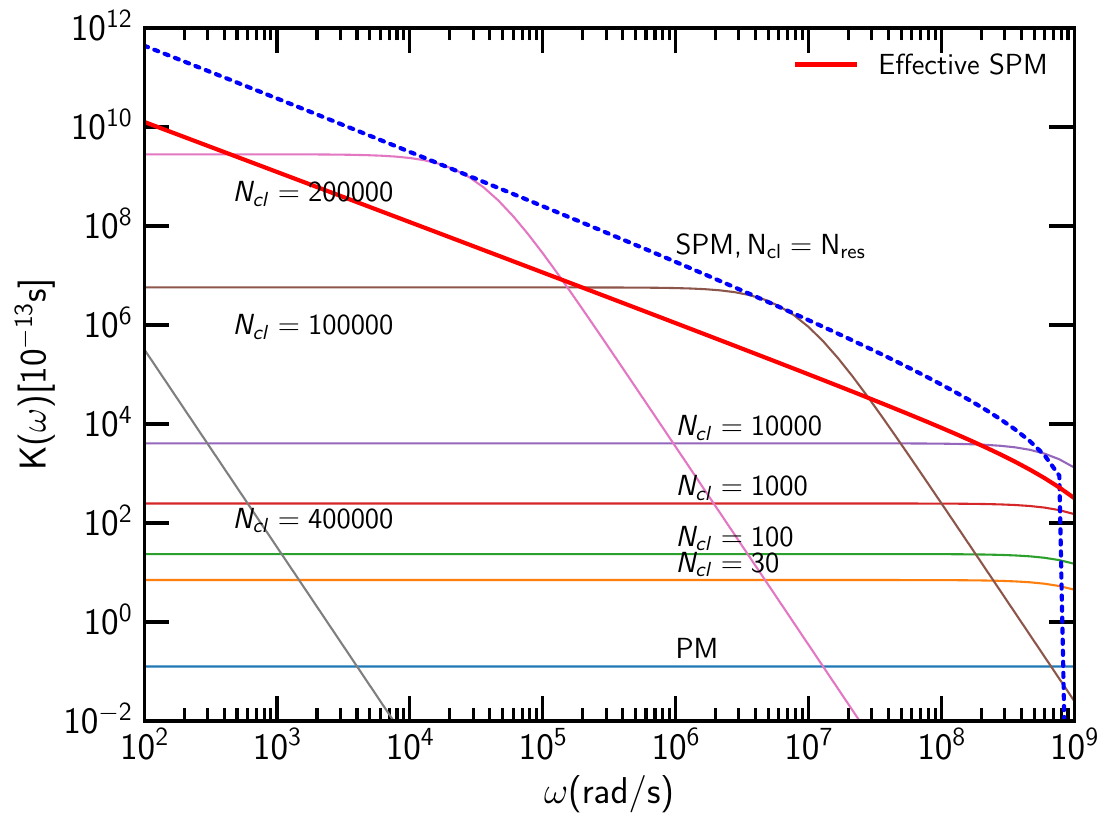}
    \put(70,60){\small \textbf{(c)} $T_{d}=200$K}
      \end{overpic}
\begin{overpic}[width=0.45\textwidth]{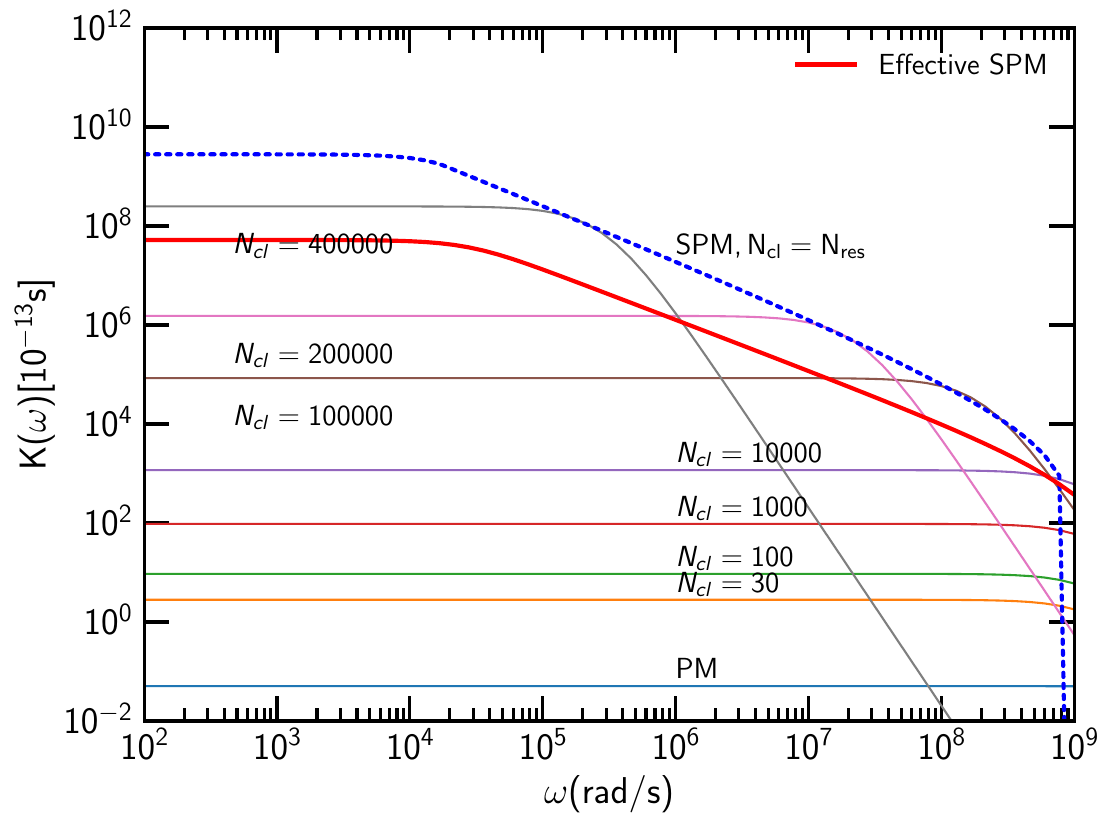}
   \put(70,60){\small \textbf{(d)} $T_{d}=500$K}
      \end{overpic}
	\caption{Magnetic dissipation strength $K=\chi_{2}(\omega,N_{\rm cl})/\omega$ as function of the grain angular frequencies for different dust temperatures from $20-500\K$. The effective strength in the thick solid line is shown for comparison.}
    \label{fig:Kappa_omega_Td}
\end{figure*}

Figure \ref{fig:Kappa_omega_Td} shows the magnetic dissipation strength $K(\omega,N_{\rm cl})$ as a function of the grain angular frequency for different dust temperatures. Same as the susceptibility, the magnetic susceptibility decreases by $100$ times when the temperature increases from the standard value of $20$K to $150$K. The transition temperature at which the susceptibility changes form increase to decrease with $T_{d}$ is determined by $\omega>\tau_{N}^{-1}$ where the denominator term becomes important. 

\section{Effects of effective superparamagnetism on grain alignment}\label{sec:alignment}
Previous studies on the effect of superparamagnetism on grain alignment focused on single-size SPIs. Here, we use the effective susceptibility of the SPI size distribution to derive the characteristic timescales and critical sizes for grain alignment. The frequency and temperature dependence of $K_{\rm eff}$ must be retained when evaluating these limits. Detailed modeling will be presented elsewhere.

\subsection{Randomization Timescale by Gas Collision and IR emission}
Evaporation of gas species from the grain surface carries away angular momentum and damps grain rotation. For our numerical estimates, we assume an oblate spheroid with semi-major axis $a$, semi-minor axis $b=sa$, and $s<1$ (e.g., \citealt{Roberge.1993}). The characteristic gas damping time is

\bea
\tau_{\gas}&=&\frac{3}{4\sqrt{\pi}}\frac{I_{\|}}{1.2n_{\rm H}m_{\rm H}v_{T}a^{4}\Gamma_{\|}}
=\frac{\sqrt{\pi}\rho sa}{3n_{\H}m_{\H}v_{T}\Gamma_{\|}}\nonumber\\
&\simeq&830\hat{\rho}\left(\frac{sa_{-5}}{n_{4}T_{\gas,1}^{1/2}\Gamma_{\|}}\right)\yr,\label{eq:tgas}
\ena
where $\hat{\rho}=\rho/(3\g\cm^{-3})$, $a_{-5}=a/(10^{-5}\cm)$, $n_{4}=n_{\H}/(10^{4}\cm^{-3})$, $T_{\gas,1}=T_{\rm gas}/(10\K)$, and $v_{T}=(2k_{B}T_{\gas}/m_{\H})^{1/2}$. The moments of inertia are $I_{\|}=8\pi\rho sa^{5}/15$ and $I_{\perp}=4\pi\rho sa^{5}(1+s^{2})/15$, the grain volume is $V_{\rm grain}=4\pi sa^{3}/3$, and $\Gamma_{\|}$ is a geometrical factor of order unity. The gas damping time also characterizes randomization by gas collisions.

Gas collisions produce a characteristic thermal rotation rate along the grain symmetry axis,
\bea
\omega_{T}&&=\left(\frac{k_{B}T_{\gas}}{I_{\|}}\right)^{1/2}\nonumber\\
&&\simeq5.2\times10^{4}\hat{\rho}^{-1/2}s^{-1/2}a_{-5}^{-5/2}T_{\gas,1}^{1/2}\s^{-1}.\label{eq:omega_th}
\ena

We define the characteristic spin rate $\omega=J/I_{\|}$ and the suprathermal parameter $\St=\omega/\omega_{T}$.

Infrared (IR) emission also damps grain rotation by carrying away angular momentum (see \citealt{Draine.1998}). For grains in thermal equilibrium, $\tau_{\rm IR}^{-1}=F_{\rm IR}\tau_{\gas}^{-1}$, where

\bea
F_{\rm IR}\simeq\frac{3.8\times10^{-3}}{a_{\eff,-5}}
\frac{U^{2/3}}{n_{4}T_{\gas,1}^{1/2}}.\label{eq:FIR}
\ena
Here $a_{\eff}=s^{1/3}a$ is the radius of a sphere of equal volume, $a_{\eff,-5}=a_{\eff}/(10^{-5}\cm)$, and $U=u_{\rm rad}/u_{\rm MMP83}$ is the radiation strength relative to the interstellar radiation field of \citet{Mathis.1983}.
The total rotational damping time, used here as $\tau_{\rm ran}$, is

\bea
\tau_{\rm ran}^{-1}=\tau_{\gas}^{-1}(1+F_{\rm IR}).\label{eq:tau_ran}
\ena
The gas-damping limits below assume $F_{\rm IR}\ll1$; when IR damping is important, the corresponding comparison with $\tau_{\rm ran}$ must retain $F_{\rm IR}$.

\subsection{Barnett Magnetic Moment and Larmor Precession}
A rotating magnetic grain acquires a Barnett magnetic moment (\citealt{Barnett.1915})

\bea
\bmu_{\rm Bar}=\frac{\chi_{\rm eff}(0)V_{\rm grain}\bOmega}{\gamma_{g}}
=-\frac{\chi_{\rm eff}(0)\hbar V_{\rm grain}}{g_{e}\mu_{B}}\bOmega,\label{eq:muBar}
\ena
where $\gamma_{g}=-g_{e}\mu_{B}/\hbar$, $g_{e}\approx2$, and $\mu_{B}=e\hbar/(2m_{e}c)\approx9.27\times10^{-21}\erg\G^{-1}$. The static effective susceptibility includes only the active SPIs, with $N_{\rm up}=\min(N_{\rm cri},N_{\rm max})$. Thus, $\chi_{\rm eff}(0)=\chi_{\rm SPM}(0,N_{\rm up})F_{\rm eff}(N_{\rm up})$.

The Larmor precession period is
\bea
\tau_{\rm Lar}=\frac{2\pi I_{\|}|\gamma_{g}|}{\chi_{\rm eff}(0)V_{\rm grain}B}
=\frac{4\pi\rho a^{2}|\gamma_{g}|}{5\chi_{\rm eff}(0)B}.\label{eq:tauB}
\ena
Following \citet{Hoangetal.2022}, the gas-damping criterion $\tau_{\rm Lar}=\tau_{\gas}$ gives the limiting size for rapid magnetic precession,
\small\begin{align}
 a_{\max,JB}^{\rm Lar}&=\frac{5\sqrt{\pi}s\chi_{\rm eff}(0)B}{12\pi|\gamma_{g}|n_{\H}m_{\H}v_{T}\Gamma_{\|}}\nonumber\\
 &\simeq1.04\times10^{7}
 \frac{N_{\rm up,5}\phi_{\rm cl,-2}F_{\rm eff,-1}\hat{p}^{2}B_{2}\hat{s}}
 {\hat{T}_{d}n_{4}T_{\gas,1}^{1/2}\Gamma_{\|}}\mum.\label{eq:amax_JB}
\end{align}

Here $N_{\rm up,5}=N_{\rm up}/10^{5}$, $\phi_{\rm cl,-2}=\phi_{\rm SPI}/0.01$, $F_{\rm eff,-1}=F_{\rm eff}(N_{\rm up})/0.1$, $B_{2}=B/(100\mu\G)$, $\hat{s}=s/0.5$, $\hat{p}=p/3$, and $\hat{T}_{d}=T_{d}/(15\K)$. The numerical coefficient is evaluated from the dimensional susceptibility in Equation (\ref{eq:chi_Ncl_zero}) with $V_{\rm Fe}=1.18\times10^{-23}\cm^{3}$. Because $N_{\rm cri}$ depends on the damping time and hence on $a$, Equation (\ref{eq:amax_JB}) is generally an implicit condition.

For example, with $T_{d}=150\K$, $T_{\gas}=100\K$, $\phi_{\rm SPI}=0.03$, $p=3$, $B=100\mu\G$, $s=0.5$, $\Gamma_{\|}=1$, $q=11/6$, $N_{\rm min}=20$, and $N_{\rm max}=5\times10^{5}$, solving this condition with $\tau_{\rm ran}=\tau_{\gas}$ gives $a_{\max,JB}^{\rm Lar}\approx0.86$ mm for $n_{\H}=10^{8}\cm^{-3}$ and $0.052\mum$ for $n_{\H}=10^{12}\cm^{-3}$. The corresponding active upper sizes are $N_{\rm up}=5\times10^{5}$ and $3.24\times10^{5}$, respectively. Once $N_{\rm up}=N_{\rm max}$, the static susceptibility and this size limit decrease as $1/T_{d}$ at fixed gas conditions. Rapid Larmor precession is a necessary timescale condition for magnetic alignment; it does not by itself determine the alignment efficiency.

\subsection{Barnett Relaxation and Internal Grain Alignment}
As a nonspherical grain wobbles, its magnetization varies in the grain frame. Dissipation associated with this variation drives internal alignment of the axis of maximum inertia with $\bJ$ (\citealt{Purcell.1979}). For an oblate grain, $h=I_{\|}/I_{\perp}=2/(1+s^{2})>1$. The frequency of this variation is $\omega_{\rm BR}=(h-1)\omega\cos\theta$, where $\theta$ is the angle between $\bJ$ and the symmetry axis; it differs from the spin frequency used for external magnetic relaxation (\citealt{Hoangetal.2022}).

The super-Barnett relaxation time is
\bea
\tau_{\rm BR}^{\rm SPM}&&=\frac{|\gamma_{g}|^{2}I_{\|}^{3}}{V_{\rm grain}K_{\rm eff}(\omega_{\rm BR})h^{2}(h-1)J^{2}}\nonumber\\
&&\simeq0.84\frac{\hat{\rho}^{2}f(s)a_{-5}^{7}}{T_{d,1}K_{\rm eff,-13}(\omega_{\rm BR})}
\left(\frac{J_{d}}{J}\right)^{2}\yr,\label{eq:tauBar_sup}
\ena
where $K_{\rm eff,-13}=K_{\rm eff}/(10^{-13}\s)$, $T_{d,1}=T_{d}/(10\K)$, $J_{d}=[I_{\|}k_{B}T_{d}/(h-1)]^{1/2}$, and
$f(s)=(s/0.5)[(1+s^{2})/1.25]^{2}$, with $f(0.5)=1$. In particular, the SPI distribution requires $K_{\rm eff}(\omega_{\rm BR})=\chi_{\rm SPM}(0,N_{\rm res})G_{\rm eff}(\omega_{\rm BR})/\omega_{\rm BR}$, evaluated at $N_{\rm res}=N_{\rm res}(\omega_{\rm BR})$. The factor $1/\omega_{\rm BR}$ must be retained when substituting this expression into the relaxation time.

Using $J=\St(I_{\|}k_{B}T_{\gas})^{1/2}$ and setting $\tau_{\rm BR}^{\rm SPM}=\tau_{\gas}$ gives

\small\begin{align}
\left[a_{\max,aJ}^{\rm BR}\right]^{6}&=
\frac{25h^{2}(h-1)\St^{2}}{16\sqrt{\pi}|\gamma_{g}|^{2}\rho}\nonumber\\
&\times\frac{K_{\rm eff}(\omega_{\rm BR})k_{B}T_{\gas}}
{n_{\H}m_{\H}v_{T}\Gamma_{\|}},\nonumber\\
 a_{\max,aJ}^{\rm BR}&\simeq0.24h^{1/3}\St^{1/3}\nonumber\\
 &\times\left[\frac{(h-1)K_{\rm eff,-13}(\omega_{\rm BR})T_{\gas,1}^{1/2}}
 {\hat{\rho}n_{4}\Gamma_{\|}}\right]^{1/6}\mum.\label{eq:amax_aJ_BR}
\end{align}

This equation must be solved with $\omega_{\rm BR}=(h-1)\St\omega_{T}(a)\cos\theta$ and the full $K_{\rm eff}(\omega_{\rm BR},T_{d})$. The apparent one-sixth-power scaling applies only when $K_{\rm eff}$ can be treated as fixed; it is not a general explicit size law for the SPI distribution. If there are multiple crossings, the largest crossing bounds the upper end of the interval with efficient relaxation.

For an illustrative gas-damping case with $T_{d}=15\K$, $T_{\gas}=10\K$, $n_{\H}=10^{4}\cm^{-3}$, $\rho=3\g\cm^{-3}$, $s=0.5$, $\Gamma_{\|}=1$, $\phi_{\rm SPI}=0.03$, $p=3$, $q=11/6$, $N_{\rm min}=20$, $N_{\rm max}=5\times10^{5}$, and $\theta=\pi/4$, numerical integration of the susceptibility gives $a_{\max,aJ}^{\rm BR}\approx69\mum$ for $\St=1$ and $132\mum$ for $\St=10$. These values are conditional on the adopted SPI distribution and the gas-damping criterion. Although a lower frequency can increase $K_{\rm eff}$, the relaxation time also depends on $J^{-2}$; the change in $K_{\rm eff}$ alone does not determine the efficiency of internal alignment.

\subsection{Magnetic Relaxation and Magnetically Enhanced Radiative Torque (MRAT) Alignment}
For rotation about the axis of maximum inertia, the magnetic relaxation time is
\bea
\tau_{\rm mag}&&=\frac{I_{\|}}{K_{\rm eff}(\omega)V_{\rm grain}B^{2}}
=\frac{2\rho a^{2}}{5K_{\rm eff}(\omega)B^{2}}\nonumber\\
&&=\frac{2\rho a_{\eff}^{2}s^{-2/3}}{5K_{\rm eff}(\omega)B^{2}}.\label{eq:tau_DG_sup}
\ena

The dimensionless magnetic relaxation strength is the ratio to the gas damping time (\citealt{HoangLaz.2016}),
\bea
\delta_{\rm mag}&&=\frac{\tau_{\gas}}{\tau_{\rm mag}}
=\frac{5\sqrt{\pi}sK_{\rm eff}(\omega)B^{2}}{6a n_{\H}m_{\H}v_{T}\Gamma_{\|}}\nonumber\\
&&\simeq0.109\frac{\hat{s}K_{\rm eff,-13}(\omega)B_{2}^{2}}{a_{-5}n_{4}T_{\gas,1}^{1/2}\Gamma_{\|}}\nonumber\\
&&\simeq1.15\times10^{8}
\frac{\hat{s}\phi_{\rm cl,-2}N_{\rm res,5}G_{\rm eff,-1}\hat{p}^{2}B_{2}^{2}}
{a_{-5}n_{4}T_{\gas,1}^{1/2}\Gamma_{\|}\hat{T}_{d}\omega_{4}}.\label{eq:delta_SPM_max}
\ena

Here $\omega_{4}=\omega/(10^{4}\s^{-1})$, $N_{\rm res,5}=N_{\rm res}(\omega)/10^{5}$, and $G_{\rm eff,-1}=G_{\rm eff}(\omega)/0.1$. For $q=11/6$, $G_{\rm eff}\approx 0.016$--$0.043$ for $10^{4}\leq\omega\leq 10^{6}\s^{-1}$ and $15\K\leq T_{d}\leq 500\K$ (Eq. \ref{eq:G_SPM}), i.e., $G_{\rm eff,-1}\approx 0.2$--$0.4$. The last line uses Equation (\ref{eq:Kappa_avg_omega_final}) and assumes $N_{\rm res}>0$. Close to either size cutoff, $G_{\rm eff}$ must be evaluated from the numerical susceptibility rather than its interior-resonance approximation. The material density $\rho$ cancels in $\delta_{\rm mag}$, and at fixed $K_{\rm eff}$ the explicit size dependence is $a^{-1}$. Its temperature and spin dependence enter through $K_{\rm eff}(\omega,T_{d})$.

For the analytical RAT model with the radiation direction parallel to the magnetic field ($\psi=0$), \citet{Hoang.2025} derived the threshold for producing high-$J$ attractors,

\bea
\delta_{\rm mag,cri}=(1+F_{\rm IR})\frac{2-q^{\rm max}}{q^{\rm max}},\label{eq:delta_mag_cri}
\ena
where $q^{\rm max}$ characterizes the RAT model (\citealt{LazHoang.2007,HoangLaz:2008gb,Herranen.2019}). We use this positive threshold for $0<q^{\rm max}<2$; for $q^{\rm max}>2$, this geometry can have high-$J$ attractors without magnetic relaxation. For a general radiation--field angle, the angular dependence of the RATs must be included.

In the IR-dominated regime, $F_{\rm IR}\gg1$, the threshold becomes
\bea
\delta_{\rm mag,cri}&&\simeq\frac{2-q^{\rm max}}{q^{\rm max}}F_{\rm IR}\nonumber\\
&&\simeq3.8\times10^{-3}\frac{2-q^{\rm max}}{q^{\rm max}}
\frac{U^{2/3}}{a_{\eff,-5}n_{4}T_{\gas,1}^{1/2}}.\label{eq:delta_mag_cri2}
\ena

Large $U$ alone does not guarantee this limit: $F_{\rm IR}$ also depends on grain size and gas conditions.

The upper MRAT threshold is determined by $\delta_{\rm mag}(T_{d},\omega)=\delta_{\rm mag,cri}$. Combining Equations (\ref{eq:delta_SPM_max}) and (\ref{eq:delta_mag_cri2}), and adopting an equilibrium temperature law $T_{d}=T_{0}U^{1/6}$ for a specified grain size, composition, and radiation spectrum, gives

\bea
\left(\frac{T_{\max}^{\rm MRAT}}{T_{0}}\right)^{4}&&\simeq22.7\frac{\hat{s}^{4/3}B_{2}^{2}}{\Gamma_{\|}}
\frac{q^{\rm max}}{2-q^{\rm max}}\nonumber\\
&&\times K_{\rm eff,-13}\!\left(\omega(T_{\max}^{\rm MRAT}),T_{\max}^{\rm MRAT}\right).\label{eq:Td_max}
\ena
Here $T_{0}$ is the equilibrium temperature at $U=1$ for the adopted grain model. This is an implicit condition because the susceptibility depends on temperature and the rotation rate can depend on the radiation strength. The explicit grain-size and gas-density factors cancel between the two relaxation strengths in this IR-dominated limit, but they can still enter through $T_{0}$ and $\omega$. A numerical upper temperature therefore requires a specified heating and rotation model and the full size-integrated susceptibility. In particular, $T_{d}\gg N_{\rm cl}T_{\rm act}$ does not make $K$ temperature independent: the Curie factor still gives $K\propto T_{d}^{-1}$ for fixed-size SPIs in their low-frequency limit.

\section{Discussion}\label{sec:discuss}
Here we discuss the various implications of the effective SPM for grain alignment and MDE.

\subsection{Effects of the SPI size distribution on effective superparmagnetic susceptibility}

We derived the effective magnetic susceptibility of dust grains containing iron clusters by assuming a power-law size distribution of superparamagnetic inclusions (SPIs). We find that accounting for the SPI size distribution fundamentally modifies the behavior of superparamagnetic susceptibility compared to models assuming a single SPI size. At zero frequency, the effective susceptibility is primarily determined by the critical blocking size,
$N_{\rm cri} = (T_d/T_{\rm act}) \ln(\nu_0 \tau_{\rm ran})$,
which corresponds to the largest SPIs that remain thermally activated over the randomization timescale. This implies that the static susceptibility is controlled by the upper end of the active size distribution rather than by a fixed characteristic inclusion size.

For rotating grains with angular velocity $\omega$, the effective response is dominated by a characteristic resonance size,
$N_{\rm res} = (T_d/T_{\rm act}) \ln(\nu_0/\omega)$,
for which the N\'eel relaxation time is equal to the rotational timescale. Both higher dust temperatures and slower grain rotation shift $N_{\rm res}$ toward larger inclusion sizes, leading to an increase in the effective susceptibility.

The frequency-dependent effective susceptibility differs qualitatively from the critically damped single-size model (Eq.~\ref{eq:chi2_cd}). Instead of exhibiting a sharp resonant feature, the SPI size distribution produces a broadened response that approaches a shallow power-law behavior at frequencies below $\nu_0$, owing to the progressive activation of larger clusters. {For the power-law distribution, $\chi_{\rm eff}(\omega)\propto[\ln(\nu_{0}/\omega)]^{2-q}$ (Eqs. \ref{eq:chi_avg_omega_final}--\ref{eq:G_SPM}). This flat spectrum requires the iron mass to be spread over a broad range of SPI sizes: for a narrow log-normal SPI size distribution, the response resembles a broadened single-size resonance near the N\'eel frequency of the SPIs at the peak of the iron mass distribution, whereas a broad log-normal distribution ($\sigma\gtrsim 0.8$ for $r_{0}=0.4$ nm) also yields a nearly flat spectrum (Appendix \ref{apdx:lognormal_spi}).}

Furthermore, in contrast to single-size SPI models where susceptibility decreases with increasing dust temperature due to enhanced thermal agitation, the effective susceptibility in the size-distribution model {with a fiducial slope of $q=11/6$} shows a slight increase with $T_d$ {until the largest SPIs ($N_{\rm max}$) become active, i.e., $T_{d}\gtrsim T_{\rm sat}\approx 110\K$ for grains at rest}. This behavior arises because higher temperatures activate larger SPIs, which contribute disproportionately to the magnetic response.

\subsection{Physical Modeling of Grain Alignment and Polarization by the MRAT Mechanism}

Superparamagnetic susceptibility plays a central role in MRAT alignment 
\citep{LazHoang.2008,HoangLaz.2016} within the radiative torque (RAT) paradigm 
\citep{Hoang.2025}. Quantifying the efficiency of MRAT alignment therefore requires an accurate evaluation of the magnetic susceptibility. In conventional single-size SPI models, the susceptibility depends on parameters such as the characteristic cluster size $N_{\rm cl}$ and the volume filling factor as $\chi_{\rm SPM}(0)\propto \boldsymbol{\phi_{\rm SPI}}N_{\rm cl}$ (see Eq.\ref{eq:chi_SPM_avg_zero1}). However, the adoption of a fixed $N_{\rm cl}$ implies the sensitive dependence of $\chi_{\rm SPM}$ on dust temperature $T_d$, since the thermal activation of SPIs depends exponentially on $N_{\rm cl}$. As a result, a single-size susceptibility model may be applicable in a limited temperature range but can significantly underestimate susceptibility in environments with higher $T_d$.

In contrast, our effective superparamagnetic susceptibility model, based on a size distribution of SPIs, naturally incorporates the temperature- and frequency-dependent activation of inclusions through the critical blocking size and the resonance size. For grains rotating at angular velocity $\omega$, the characteristic resonance size
$ N_{\rm res} = (T_d/T_{\rm act}) \ln(\nu_0/\omega)$
determines the dominant contribution to the magnetic response. 

Within the standard RAT framework, grain alignment occurs at low-$J$ and high-$J$ attractor points with angular velocities $\omega_{\rm low-J} \sim \omega_T$ and $\omega_{\rm high-J} = \omega_{\rm RAT}$ \citep{Hoangetal.2022}. Since $\omega_{\rm low-J} < \omega_{\rm high-J}$, the resonance condition implies
$
N_{\rm res}^{\rm low-J} > N_{\rm res}^{\rm high-J}
$
for a given radiation field and dust temperature. Consequently, low-$J$ grains activate larger SPIs and can experience enhanced magnetic relaxation relative to high-$J$ grains. This coupling between grain rotation rate, dust temperature, and inclusion size distribution introduces a physically motivated variation in alignment efficiency across environments.

In future work, we will incorporate the size-distribution–based superparamagnetism model into our publicly available codes \textsc{Polaris} \citep{Reissl.2016,Gianghoang.2023} and \textsc{DustPol-py} \citep{LeeHoang.2020,TramHoang:2021} to predict polarization signatures under specific astrophysical conditions and to compare directly with observational data.

\subsection{Towards Constraining Iron Inclusions in Dust with Polarization}

Constraining the abundance of iron inclusions in interstellar dust is crucial for understanding the cosmic iron budget and the magnetic properties of dust grains. Polarization observations provide a promising avenue for such constraints, since both grain alignment efficiency and dust polarization depend sensitively on the magnetic susceptibility determined by iron inclusions, as demonstrated in our previous studies \citep{Giangetal.2024,Gianghoang.2023}.

In conventional single-size SPI models, however, there exists a degeneracy between the characteristic cluster size $N_{\rm cl}$ and the volume filling factor $\phi_{\rm SPI}$, as different combinations of these parameters can yield similar magnetic susceptibilities $\chi_{\rm SPM}(0)$. This degeneracy limits the ability of polarization measurements to uniquely constrain the iron content in dust.

In contrast, our size-distribution–based superparamagnetism model naturally determines the effective magnetic response through the critical blocking size ($N_{\rm cri}$) and the resonance size ($N_{\rm res}$), both of which are governed by dust temperature and grain rotation rate. As a result, the effective susceptibility depends primarily on the volume filling factor $\phi_{\rm SPI}$, reducing the degeneracy with cluster size. This framework therefore provides a physically motivated pathway to constrain the abundance of iron inclusions using multi-wavelength polarization observations.

\subsection{Implications for Magnetic Dipole Emission Spectrum}

Thermal fluctuations of magnetic dipoles produce magnetic dipole emission (MDE) of emissivity $j_{\nu}\propto  \mu_{2}(\omega/2\pi)B_{\nu}(T_{d})$ where $\mu_{2}=1+4\pi \chi_{2}$ and $B_{\nu}(T_{d})$ is the Planck function \citep{DraineLaz.1999}. 

In the DL99 model, the magnetic susceptibility was evaluated for uniform-sized SPIs, resulting in a resonant spectrum characterized by a peak at the N\'eel relaxation frequency of the inclusions. For a single inclusion size, the susceptibility exhibits a narrow resonance at a characteristic frequency $\omega_{\rm res} \sim \tau_{N}^{-1}$, which depends exponentially on cluster size and dust temperature through the N\'eel relaxation time. The susceptibility decreases rapidly away from this resonance. When a power-law distribution of SPI sizes is considered, the integration over a continuous ensemble of inclusion sizes produces a superposition of individual resonant peaks. This superposition smooths the spectral response and enhances the effective susceptibility at frequencies below the characteristic attempt frequency $\nu_0$. As a result, instead of a sharply peaked resonance, the effective susceptibility exhibits a broadened response that approaches a power-law–like spectrum at low frequencies. A detailed modeling of magnetic dipole emission based on the new effective susceptibility will be presented in future work.



\section{Summary}\label{sec:summary}
We have studied the effect of a size distribution of superparamagnetic inclusions (SPIs) on the magnetic susceptibility of dust grains containing iron clusters. Our main results are summarized as follows:
\begin{itemize}
    \item We derived the effective susceptibility by integrating over a power-law size distribution of SPIs, $dn/dN\sim N^{-q}$. The zero-frequency effective susceptibility is primarily determined by the critical blocking size, $N_{\rm cri}$, corresponding to the largest SPI size that remains superparamagnetic under thermal fluctuations, reduced by a factor $F_{\rm eff}\sim 0.1$. {Above $T_{d}\approx 110\K$, $N_{\rm cri}$ exceeds the maximum SPI size $N_{\rm max}$, all SPIs are active, and $\chi_{\rm eff}(0)\propto 1/T_{d}$.}
    
    \item For a grain rotating at angular frequency $\omega$, the effective magnetic susceptibility is dominated by a characteristic resonance size, $N_{\rm res}$, defined by the condition that the N\'eel relaxation rate $\tau_{N}^{-1}$ equal the grain angular frequency. The resonance size depends on the grain rotation rate and dust temperature as $N_{\rm res}=(T_{d}/T_{\rm act})\ln(\nu_{0}/\omega)$. 
    
    \item The effective susceptibility of grains with SPIs can be described by the susceptibility of SPIs at the resonance size and a reduction factor {$G_{\rm eff}\simeq (\pi/2)(2-q)(N_{\rm res}/N_{\rm max})^{2-q}/\ln(\nu_{0}/\omega)\approx 0.02$--$0.04$ for $q=11/6$}. It slightly changes with the angular frequency below $\nu_{0}${, as $[\ln(\nu_{0}/\omega)]^{2-q}$,} in contrast to the peaky profile for the case of single-size SPIs, due to the activation of larger iron clusters at lower frequencies.

    \item {For a fiducial SPI size-distribution slope of $q=11/6$}, the effective susceptibility increases slightly with dust temperature, which is opposite to the rapid decrease as $\exp\left(\rm N_{\rm cl}T_{\rm act}/T_{d}\right)/T_{d}$ predicted for single-size SPIs{, until the largest SPIs are activated}. This trend arises from the thermal activation of larger inclusions at higher temperatures.

    \item {For a narrow log-normal SPI size distribution, the zero-frequency susceptibility approximately follows Curie's law with an effective size set by the iron mass-weighted mean SPI size, and the frequency-dependent susceptibility resembles a broadened single-size resonance near the N\'eel frequency of the SPIs at the peak of the iron mass distribution, rather than the flat spectrum of the power-law distribution. A broad log-normal distribution ($\sigma\gtrsim 0.8$ for $r_{0}=0.4$ nm), in which the iron mass is spread over a wide range of SPI sizes, behaves like the power law.}
    
    \item The effective susceptibility model enables us to predict the grain alignment efficiency with ambient magnetic fields within the MRAT mechanism through the critical blocking size and resonance SPI size for the different grain rotation rates and radiation fields. 
    
    \item The effective susceptibility model is expected to yield a power-law spectrum of magnetic dipole emission because the latter depends on the imaginary part of the magnetic susceptibility, which would be radically different from that derived under the assumption of uniform-sized SPIs.
    
\end{itemize}

\section*{Acknowledgements}
This work was supported by the major research project (No. 2026183200) from Korea Astronomy and Space Science Institute (KASI) funded by the Ministry of Science and ICT (MSIT). This work was partially funded by a grant from the Simons Foundation (SFI-MPS-T-Institutes-00021894, HTN). We thank the ICISE staff for excellent support. {We thank the anonymous referee for comments that helped improve the manuscript.}


\bibliography{ms.bbl}

\begin{appendix}
    
\section{Size distribution of SPIs and effects on effective susceptibility}\label{apdx:Cconst}
\subsection{Power-law distribution}
The size distribution of SPIs is uncertain. In our study, we assumed that they follow the power-law distribution of $dn/dN \propto N^{-q}$. {Since $dn/dN$ is the number density of SPIs per unit $N$ and per unit physical volume of the host grain, the total volume of SPIs inside a grain of volume $V_{\rm grain}$ is calculated as}
\bea
V_{\rm SPI} &&=V_{\rm grain}\int_{N_{\rm min}}^{N_{\rm max}}(dn/dN)NV_{\rm Fe}dN\nonumber\\
&&=CV_{\rm grain}V_{\rm Fe}\int_{N_{\rm min}}^{N_{\rm max}}N^{1-q}dN\nonumber\\
&&=\frac{CV_{\rm grain}V_{\rm Fe}N_{\rm max}^{2-q}}{2-q}\left[1-(N_{\rm min}/N_{\rm max})^{2-q} \right].
\ena
where $ NV_{\rm Fe}$ is the volume of each cluster, and $V_{\rm Fe}=1.18 \times 10^{-23}\cm^{3}$ is the atomic Fe volume. {Using the definition of the volume filling factor of iron inclusions in the grain, $\phi_{\rm SPI}=V_{\rm SPI}/V_{\rm grain}$, the grain volume cancels, and one obtains}
\bea
C= \frac{(2-q)\phi_{\rm SPI}}{V_{\rm Fe}\left[N_{\rm max}^{2-q}-N_{\rm min}^{2-q}\right]}.\label{eq:Cnorm1}
\ena

The slope of SPI size distribution $q$ is uncertain. Here, we discuss the relationship between this power-law SPI distribution with the grain size distribution of interstellar dust. The size distribution of interstellar dust is usually described by a power-law
\bea
\frac{dn}{dr} \propto r^{-\alpha},
\ena
where $\alpha=3.5$ is the standard MRN distribution \cite{Mathis.1977}.

{If we assume that the radii $r$ of the SPIs themselves (not those of their host grains) follow a similar power law, $dn/dr\propto r^{-\alpha}$, then one can obtain the link between the slope $q$ of the SPI size distribution and $\alpha$. We note that this assumption is not constrained by observations or laboratory data and needs to be tested.}

Because $N_{\rm cl}\sim r^{3}$, one obtains
\bea
\frac{dn}{dN}&&= \left(\frac{dn}{dr}\right)\left(\frac{dr}{dN}\right),\nonumber\\
&&\propto r^{-(\alpha+2)}\sim N_{\rm cl}^{-(\alpha+2)/3}
\ena
which yields $q=(\alpha+2)/3$. If SPIs follow the standard MRN distribution slope of $\alpha=3.5$, then $q=11/6<2$.

For the typical MRN slope of $q=11/6$, the total volume of SPIs is determined largest SPIs. However, if SPIs have a steeper slope of $q>2$, the SPI mass is dominated by smallest clusters of $N_{\rm min}$.

The slope of SPI size distribution also has important implications for the effective susceptibility. From Equation (\ref{eq:chi_avg_zero}), one can see that for the slope of $q<3$ ($\alpha<7$), $\chi_{\rm eff}(0) $ is dominated by large SPIs, but for the rather steep slope of $q>3 (\alpha>7)$, it is dominated by smallest SPIs of size $N_{\rm min}$. However, the scenario with such an extreme steep slope of the SPI radius distribution of $\alpha>7$ is unlikely because due to grain growth and destruction in the ISM.

\subsection{Log-normal SPI size distribution}\label{apdx:lognormal_spi}
Since the SPI size distribution is not well constrained, for sake of completeness, we also consider a log-normal size distribution of SPIs (see \citealt{HoangLaz.2016b}),
\bea
\frac{dn}{dr}=\frac{B}{r}\exp\left[-\frac{\ln^{2}(r/r_{0})}{2\sigma^{2}}\right],
\ena
where $dn/dr$ is the SPI number density per unit radius and host-grain volume, $r_{0}$ is the median radius, and $\sigma$ is the logarithmic width. The normalization $B$ (in units of $\cm^{-3}$) is determined by the volume filling factor,
\bea
\int_{r_{\min}}^{r_{\max}}\frac{4\pi r^{3}}{3}\frac{dn}{dr}dr=\phi_{\rm SPI}.\label{eq:Bdnda}
\ena

For compact SPIs, $N\propto r^{3}$. Adopting $N=20(r/0.35\,{\rm nm})^{3}$ (see \citealt{Billas:1994p6378}), the limits $N_{\rm min}=20$ and $N_{\rm max}=5\times10^{5}$ correspond to $r_{\rm min}=0.35$ nm and $r_{\rm max}\approx10$ nm. The distribution in $N$ is also log-normal, with median $N_{0}=20(r_{0}/0.35\,{\rm nm})^{3}$ and width $\sigma_{N}=3\sigma$. The iron mass per logarithmic size interval peaks at $r_{p}=r_{0}e^{3\sigma^{2}}$ \citep{Li:2001p4761}, or $N_{p}=N_{0}e^{9\sigma^{2}}$.

The effective susceptibility is calculated using Equations (\ref{eq:chi_SPM_avg_zero1}) and (\ref{eq:chi_SPM_omega_avg}), with the log-normal distribution normalized over $[N_{\rm min},N_{\rm max}]$. When all SPIs are active, $\chi_{\rm eff}(0)=\chi_{\rm SPM}(0,N_{\rm eff})$, where $N_{\rm eff}=\langle N^{2}\rangle/\langle N\rangle$ is the iron-mass-weighted mean size. For negligible truncation, $N_{\rm eff}=N_{p}e^{\sigma_{N}^{2}/2}=N_{0}e^{27\sigma^{2}/2}$. Thus, the zero-frequency susceptibility is determined by $N_{\rm eff}$ rather than by the median size $N_{0}$.

\begin{figure*}
\centering
\includegraphics[width=0.45\textwidth]{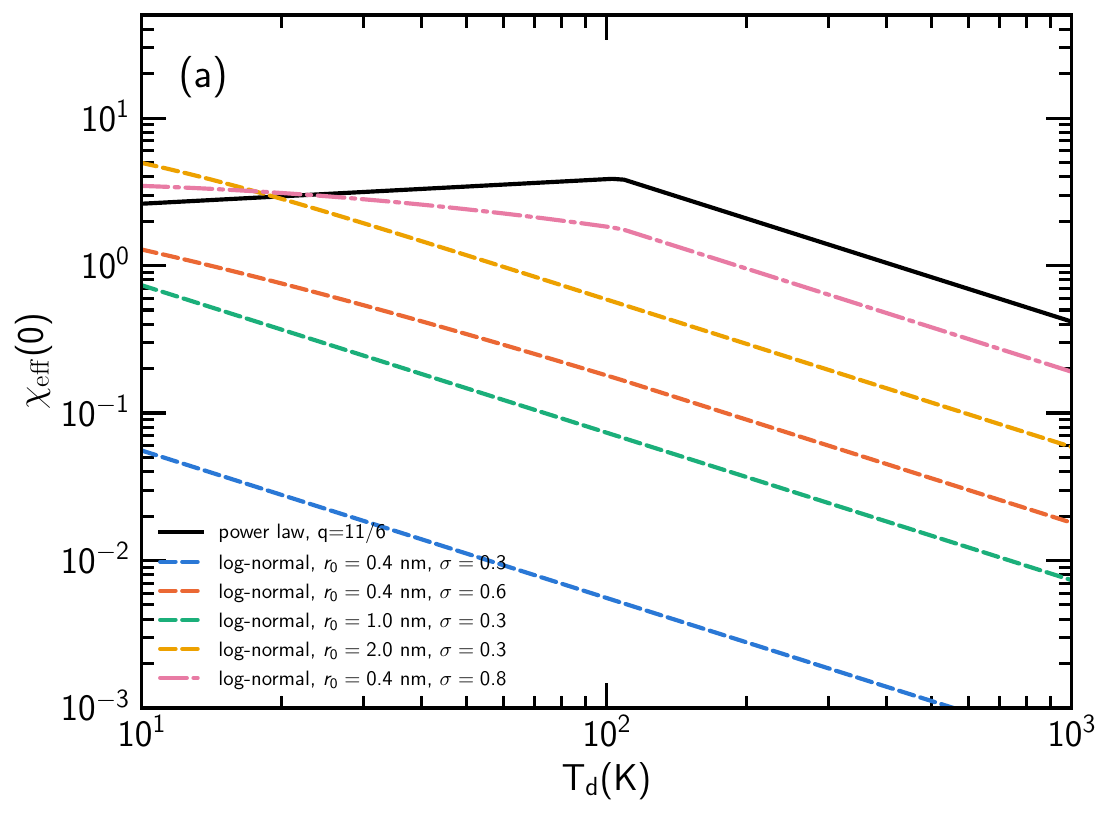}
\includegraphics[width=0.45\textwidth]{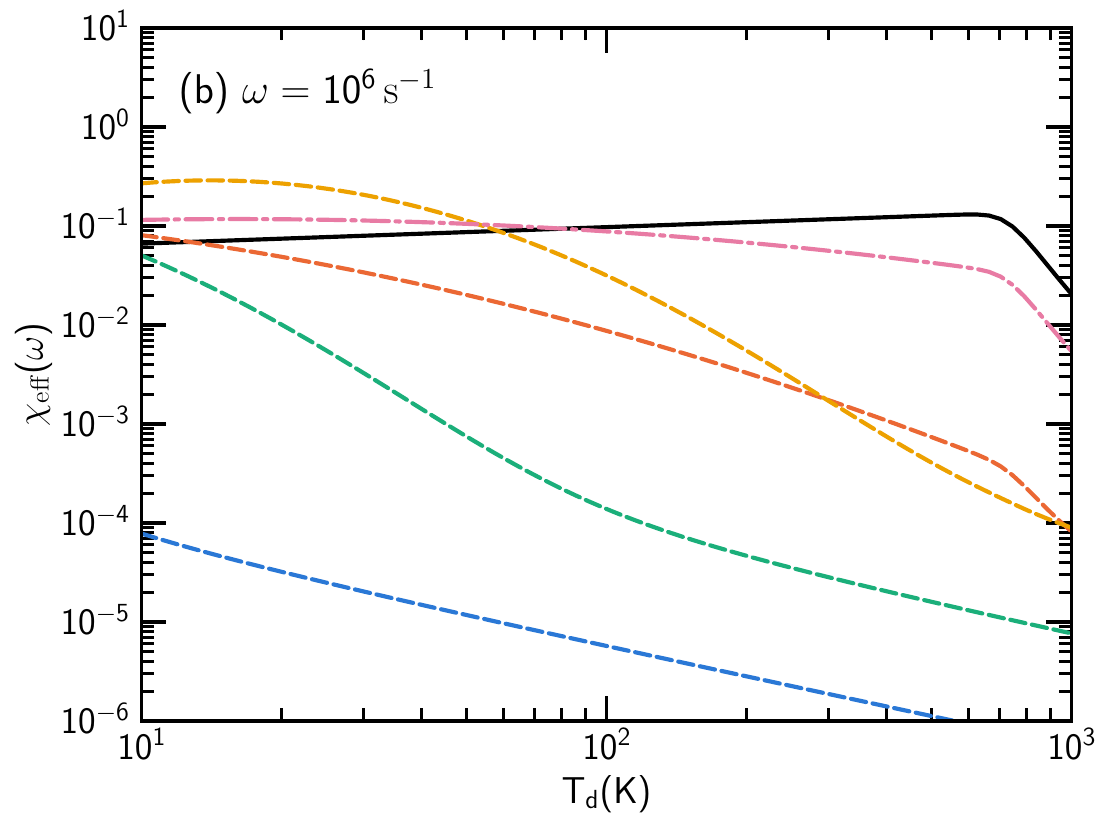}
\includegraphics[width=0.45\textwidth]{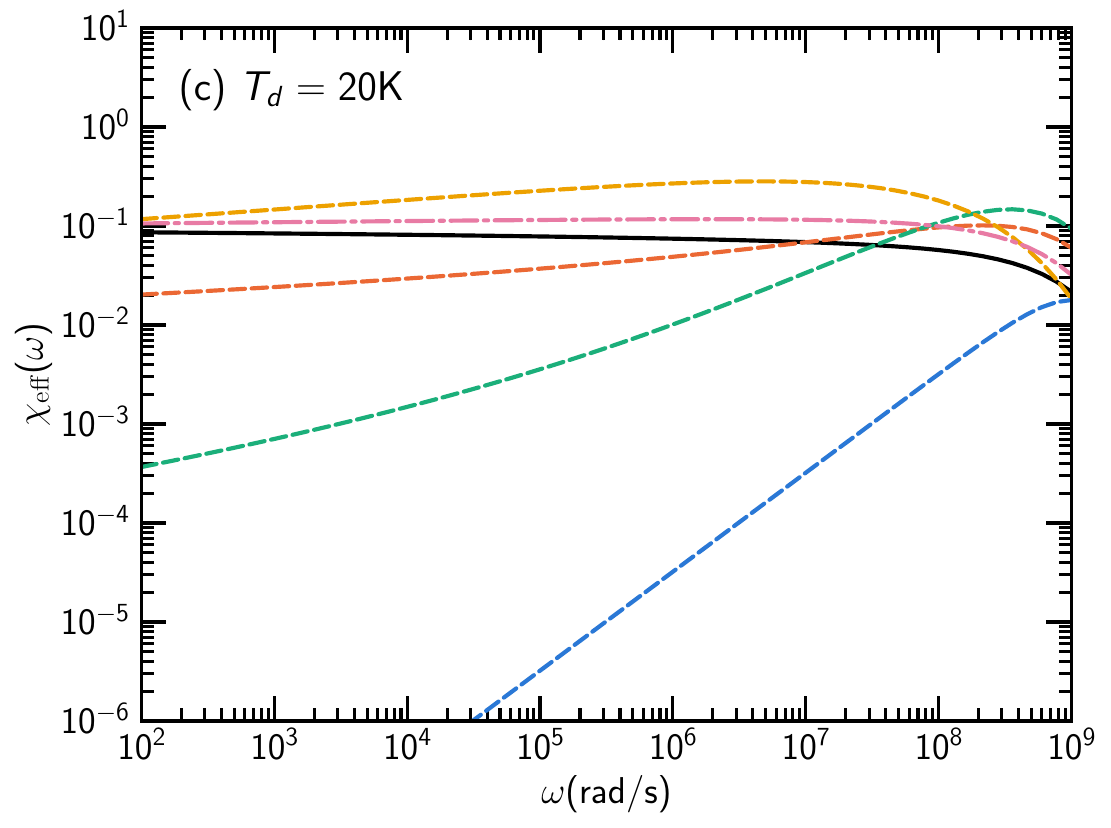}
\includegraphics[width=0.45\textwidth]{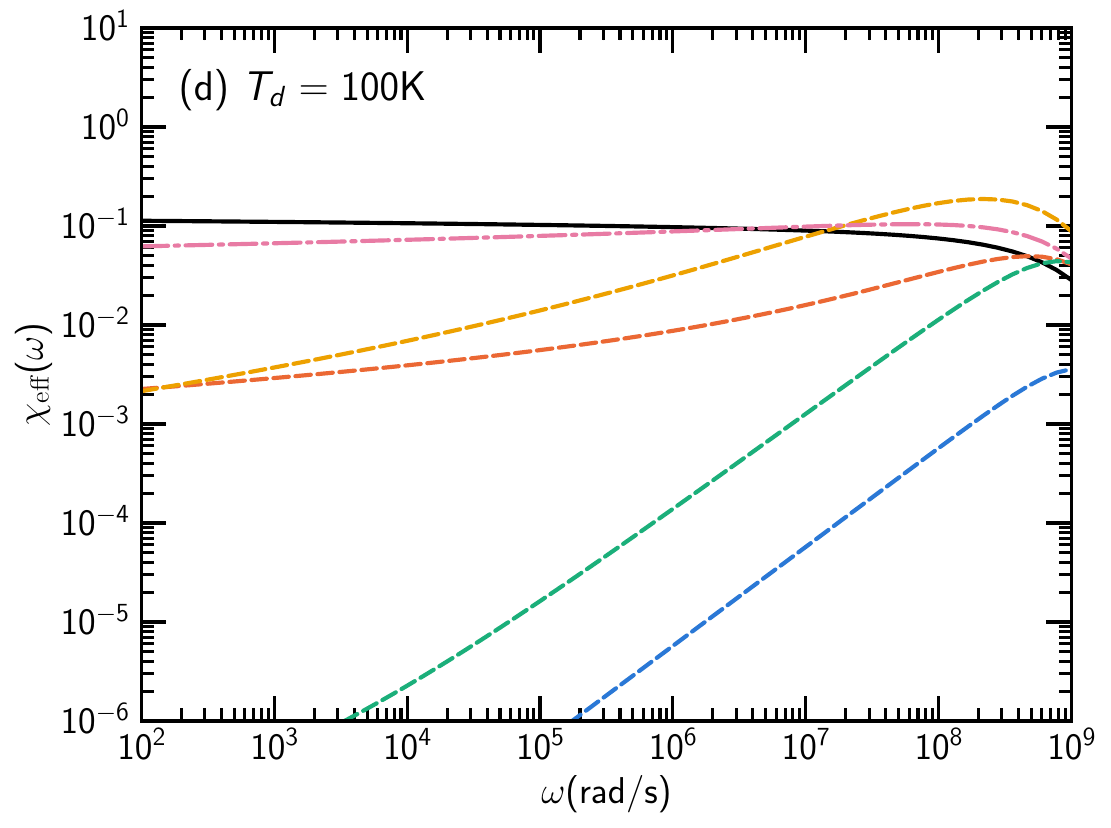}
\caption{Effective susceptibility for log-normal SPI distributions with $(r_{0},\sigma)=(0.4\,{\rm nm},0.3)$, $(0.4\,{\rm nm},0.6)$, $(1\,{\rm nm},0.3)$, and $(2\,{\rm nm},0.3)$ (dashed lines), and $(0.4\,{\rm nm},0.8)$ (dash-dotted line), compared with the power-law distribution of $q=11/6$ (solid line). All models have $\phi_{\rm SPI}=0.03$. (a) $\chi_{\rm eff}(0)$ vs. $T_{d}$, with SPIs larger than $N_{\rm cri}$ blocked. (b) $\chi_{\rm eff}(\omega)$ vs. $T_{d}$ for $\omega=10^{6}\s^{-1}$. (c) and (d) $\chi_{\rm eff}(\omega)$ vs. $\omega$ for $T_{d}=20\K$ and $100\K$.}
\label{fig:chi_eff_lognorm}
\end{figure*}

Figure \ref{fig:chi_eff_lognorm} compares five log-normal models with the power-law distribution for the same iron content. The four narrower models have $N_{\rm eff}\approx10^{2}$, $3.9\times10^{3}$, $1.6\times10^{3}$, and $1.3\times10^{4}$, respectively, including the finite size limits. Most SPIs are active at $T_{d}\gtrsim50$--$100\K$, so $\chi_{\rm eff}(0)$ follows Curie's law, $\chi_{\rm eff}(0)\propto1/T_{d}$. At $T_{d}=10\K$, blocking reduces the susceptibility of the $(0.4\,{\rm nm},0.6)$ and $(2\,{\rm nm},0.3)$ models by $\approx30\%$ and $15\%$, respectively, relative to the unblocked values (panel (a)).

For rotating grains, the narrower distributions produce a broadened single-size resonance near the N\'eel frequency of SPIs at the mass peak, $\omega_{p}=\nu_{0}\exp(-N_{p}T_{\rm act}/T_{d})$. Their susceptibility can exceed that of the power-law distribution near the resonance, but is much lower at low frequencies and high temperatures. At a fixed $\omega$, it decreases steeply once $N_{\rm res}\gg N_{p}$ (panels (b)--(d)).

The broad model, $(r_{0},\sigma)=(0.4\,{\rm nm},0.8)$, has $N_{p}\approx9.5\times10^{3}$, $\sigma_{N}=2.4$, and $N_{\rm eff}\approx4.1\times10^{4}$. Its large-size tail is partly blocked below $T_{\rm sat}$, so $\chi_{\rm eff}(0)$ decreases more slowly than $1/T_{d}$ and follows Curie's law only above $T_{\rm sat}$. For rotating grains, this model gives a nearly flat spectrum, varying by less than a factor of 2 over $\omega=10^{2}$--$10^{8}\s^{-1}$ at $T_{d}=20$ and $100\K$, and by a factor of $\approx3$ over $T_{d}=10$--$600\K$ at $\omega=10^{6}\s^{-1}$.

This behavior follows from $N_{\rm res}\propto\ln(\nu_{0}/\omega)$ (Eq. \ref{eq:Npeak}): the resonant sizes span only a factor of $\approx7$ over $\omega=10^{2}$--$10^{8}\s^{-1}$. A nearly flat spectrum is obtained when the iron mass per logarithmic size interval varies slowly over this range, as for the power law and the broad log-normal model. Thus, the flat spectrum does not require a power-law SPI distribution.

\end{appendix}

\end{document}